\documentclass[showkeys,showpacs,superscriptaddress,nofootinbib]{revtex4-2}
\usepackage{amsmath}
\usepackage{amssymb}
\usepackage{graphicx}
\usepackage{color}

\begin{document}

\title{Dirac stars with wormhole topology
}
\author{
Vladimir Dzhunushaliev
}
\email{v.dzhunushaliev@gmail.com}
\affiliation{
Institute of Nuclear Physics, Almaty 050032, Kazakhstan
}
\affiliation{
Department of Theoretical and Nuclear Physics,  Al-Farabi Kazakh National University, Almaty 050040, Kazakhstan
}
\affiliation{
Academician J.~Jeenbaev Institute of Physics of the NAS of the Kyrgyz Republic, 265 a, Chui Street, Bishkek 720071, Kyrgyzstan
}
\affiliation{
International Laboratory for Theoretical Cosmology, Tomsk State University of Control
Systems and Radioelectronics (TUSUR),
Tomsk 634050, Russia
}

\author{Vladimir Folomeev}
\email{vfolomeev@mail.ru}
\affiliation{
Institute of Nuclear Physics, Almaty 050032, Kazakhstan
}
\affiliation{
Academician J.~Jeenbaev Institute of Physics of the NAS of the Kyrgyz Republic,
265 a, Chui Street, Bishkek 720071, Kyrgyzstan
}
\affiliation{
International Laboratory for Theoretical Cosmology, Tomsk State University of Control
Systems and Radioelectronics (TUSUR),
Tomsk 634050, Russia
}

\author{
Sayabek Sakhiyev
}
\email{s.sakhi@inp.kz}
\affiliation{
Institute of Nuclear Physics, Almaty 050032, Kazakhstan
}


\begin{abstract}
We consider configurations consisting of a gravitating nonlinear spinor field and a massless ghost scalar field 
providing a nontrivial spacetime topology. For such a mixed system, we have constructed families of
asymptotically flat asymmetric solutions describing Dirac stars with wormhole topology. The physical properties 
of such systems are completely determined by the values of three input quantities: the nonlinearity parameter of the spinor field, the spinor frequency, 
and the throat parameter. Depending on the specific values of these parameters, the configurations may be regular or singular and possess one 
or two throats, as well as they may contain possible horizons of different kinds.
Furthermore, because of the asymmetry, masses and sizes of the configurations observed at the two asymptotic ends of the wormhole 
may differ considerably. For large negative values of the nonlinearity parameter, 
it is possible to obtain configurations whose masses are comparable to the Chandrasekhar mass
and effective radii are of the order of kilometers.
\end{abstract}

\pacs{04.40.Dg, 04.40.--b, 04.40.Nr}

\keywords{Dirac stars, nonlinear spinor fields, nontrivial topology, compact gravitating configurations}

\maketitle

\section{Introduction}

Various compact strongly gravitating objects (white dwarfs, neutron and boson stars, black holes) are of great interest both with theoretical and with observational
point of view. Due to the presence of a strong gravitational field and high densities of matter  supporting configurations of such type, 
their physical characteristics differ considerably from those of ordinary stars, whose matter is confined in equilibrium by a weak gravitational field.
Modeling of such  strongly gravitating systems is usually carried out either by applying some effective hydrodynamical description of matter (an equation of state) or
by using various fundamental fields. In the latter case, most commonly one uses scalar (boson) fields with spin 0; this enables one to get objects with sizes varying 
from atoms to stars~\cite{Schunck:2003kk,Liebling:2012fv}. However, this does not preclude the possibility of constructing compact objects consisting of
fields with higher-order spins. For example, these can be compact, 
strongly gravitating configurations consisting of massless spin-1 vector fields (Yang-Mills systems~\cite{Bartnik:1988am,Volkov:1998cc}) or
of massive vector fields (Proca stars~\cite{Brito:2015pxa,Herdeiro:2017fhv,Sanchis-Gual:2019ljs}). 

In turn, gravitating fractional-spin fields can also form various compact objects.
In the case of spin-$1/2$ fields, there are spherically and axially symmetric configurations supported by both linear
 \cite{Finster:1998ws,Finster:1998ux,Herdeiro:2017fhv,Herdeiro:2019mbz,Herdeiro:2021jgc} and nonlinear spinor 
 fields~\cite{Krechet:2014nda,Adanhounme:2012cm,Dzhunushaliev:2018jhj,Bronnikov:2019nqa,Dzhunushaliev:2019kiy,Dzhunushaliev:2019uft}.
Nonlinear spinor fields can also be used for constructing  cylindrically symmetric solutions~\cite{Bronnikov:2004uu}, including wormholes~\cite{Bronnikov:2009na}, as well as
in studying various cosmological problems (see Refs.~\cite{Ribas:2010zj,Ribas:2016ulz,Saha:2016cbu} and references inside).

Usually, in studying gravitating configurations, it is assumed that they possess a trivial spacetime topology. However, one can also imagine a situation 
when, apart from ordinary matter, a system contains some exotic matter providing the possibility of emerging  
a nontrivial wormhole-like topology. As such exotic matter, one might take, for example, one of the forms of dark energy, which, according to the modern view, ensures the 
acceleration of the present Universe~\cite{AmenTsu2010}. The observational data indicate~\cite{Ade:2015xua}
that the possibility of the existence in the Universe of dark matter violating the null/weak energy condition is not also excluded.

If such type of exotic matter does really exist in the present Universe, 
it is natural to suppose that localized wormhole-type objects consisting of such matter may also exist.
Indeed, the fact that the exotic matter violates the null/weak energy condition can provide the possibility for the presence of
nontrivial topology inherent in a wormhole.
In the simplest case, the exotic matter can be described by
the so-called ghost (or phantom) scalar fields, which may be massless~\cite{Bronnikov:1973fh,Ellis:1973yv,Ellis:1979bh}
or have a potential energy~\cite{Kodama:1978dw,Kodama:1978zg}.

However, apart from pure wormhole systems containing only exotic matter, it is also possible to consider objects in which exotic matter is only one of the components,
and other types of matter are present as well. Such mixed configurations with nontrivial topology can be exemplified by the
systems considered by us earlier, where, along with a ghost scalar field, there
appears either neutron matter~\cite{Dzhunushaliev:2011xx,Dzhunushaliev:2012ke,Dzhunushaliev:2013lna,Dzhunushaliev:2014mza,Aringazin:2014rva,Dzhunushaliev:2015sla,Dzhunushaliev:2016ylj},
 or an ordinary scalar field~\cite{Dzhunushaliev:2014bya}, or a chiral field~\cite{Charalampidis:2013ixa}.
  In turn, such mixed configurations can also be generalized to systems with rotation~\cite{Hoffmann:2017vkf,Hoffmann:2018oml}.
  
In the present paper we  study mixed systems consisting of a ghost scalar field
(which provides a nontrivial topology) and a nonlinear Dirac field.  As mentioned
above, in Refs.~\cite{Dzhunushaliev:2018jhj,Dzhunushaliev:2019kiy,Dzhunushaliev:2019uft},
we have examined Dirac stars created by nonlinear spinor fields and having trivial spacetime topology.
Due to the presence of the nonlinearity, such systems possess a number of properties interesting from the point of view of observational astrophysics: their masses and sizes may be comparable to characteristics
typical of neutron stars. The purpose of the present work is to study the influence that the presence of
nontrivial topology has on the structure of the corresponding solutions and characteristics of the resulting mixed Dirac-star-plus-wormhole systems.
 In this connection notice that Ref.~\cite{Hao:2023igi} studies a similar system supported by a linear spinor field. 
 In the present paper, we generalize those solutions to the case of a nonlinear Dirac field and
revisit the solutions of Ref.~\cite{Hao:2023igi} for the linear case.

In connection with the possibility of obtaining regular wormhole-like solutions with spinor fields, it is of interest to mention  static, {\it symmetric} with respect to the wormhole throat solutions 
obtained within Einstein-Dirac-Maxwell  theory~\cite{Blazquez-Salcedo:2020czn} 
and their criticism in Refs.~\cite{Bolokhov:2021fil,Danielson:2021aor,Konoplya:2021hsm}. As a possibility to avoid the drawbacks of the model of Ref.~\cite{Blazquez-Salcedo:2020czn},
the work~\cite{Konoplya:2021hsm} suggests static {\it asymmetric} wormholes supported
by smooth metric and matter fields. However, the study of the time-dependent Einstein-Dirac-Maxwell model performed in Ref.~\cite{Kain:2023ann}
revealed that, in the course of time, a black hole develops in such a system, and correspondingly such wormholes are not traversable.

Notice here that in the present paper we consider a system involving a classical spinor field. 
In doing so, it is  assumed that the latter represents a set of four complex-valued spacetime functions which transform according to the 
spinor representation of the Lorentz group. Of course, realistic particles with spin 1/2 must be described by quantum spinor fields.
Nevertheless,  classical spinors may be considered as appearing when one uses  some effective description 
of more complex quantum systems, see Ref.~\cite{ArmendarizPicon:2003qk}.

The paper is organized as follows. In Sec.~\ref{prob_statem}, we give the general-relativistic equations for the configurations under consideration.
These equations are solved numerically in Sec.~\ref{num_sol} for the linear spinor field
(Sec.~\ref{linear_field}) and for the nonlinear spinor field (Sec.~\ref{nonlinear_field}). 
Then we consider some astrophysical applications of the systems under consideration in Sec.~\ref{limit_conf} and discuss the question of stability in Sec.~\ref{stability}. 
Finally, in Sec.~\ref{concl}, we summarize and discuss the results obtained.

\section{Statement of the problem and general equations}
\label{prob_statem}

Within Einstein's general relativity, we consider compact configurations supported by a spinor field minimally coupled to
 a massless ghost scalar field. The total action for such a system can be written in the form
[we use the metric signature $(+,-,-,-)$ and natural units $c=\hbar=1$]
\begin{equation}
\label{action_gen}
	S_{\text{tot}} = - \frac{1}{16\pi G}\int d^4 x
		\sqrt{-g} R +S_{\text{sp}}+S_{\text{sf}} ,
\end{equation}
where $G$ is the Newtonian gravitational constant and $R$ is the scalar curvature.  

The action  $S_{\text{sp}}$ for the spinor field $\psi$ appearing in Eq.~\eqref{action_gen} can be obtained from the Lagrangian 
\begin{equation}
	L_{\text{sp}} =	\frac{\imath}{2} \left(
			\bar \psi \gamma^\mu \psi_{; \mu} -
			\bar \psi_{; \mu} \gamma^\mu \psi
		\right) - \mu \bar \psi \psi - F(S),
\label{lagr_sp}
\end{equation}
where $\mu$ is the mass of the spinor field and
the semicolon denotes the covariant derivative defined as
$
\psi_{; \mu} =  [\partial_{ \mu} +1/8\, \omega_{a b \mu}\left( \gamma^a  \gamma^b- \gamma^b  \gamma^a\right)]\psi
$.
Here $\gamma^a$ are the Dirac matrices in the standard representation in flat space
 [see, e.g.,  Ref.~\cite{Lawrie2002}, Eq.~(7.27)]. In turn, the Dirac matrices in curved space,
$\gamma^\mu = e_a^{\phantom{a} \mu} \gamma^a$, are derived  using the tetrad
 $ e_a^{\phantom{a} \mu}$, and $\omega_{a b \mu}$ is the spin connection
[for its definition, see Ref.~\cite{Lawrie2002}, Eq.~(7.135)].
The Lagrangian~\eqref{lagr_sp} involves an arbitrary nonlinear term $F(S)$ with the invariant $S$ that can depend on
$
	\left( \bar\psi \psi \right),
	\left( \bar\psi \gamma^\mu \psi \right)
	\left( \bar\psi \gamma_\mu \psi \right)$, or
	$\left( \bar\psi \gamma^5 \gamma^\mu \psi \right)
	\left( \bar\psi \gamma^5 \gamma_\mu \psi \right)$.

The action for the real ghost scalar field $S_{\text{sf}}$ appearing in \eqref{action_gen} can be found from the Lagrangian
$$	L_{\text{sf}} =-\frac{1}{2}\partial_\mu\phi \partial^\mu\phi.
$$

Then, by varying the action \eqref{action_gen} with respect to the metric, to the spinor and scalar fields, we obtain the Einstein, Dirac, and scalar field equations in curved spacetime:
\begin{eqnarray}
G_\mu^\nu\equiv 	R_{\mu}^\nu - \frac{1}{2} \delta_{\mu }^\nu R &=&
	8\pi G \,T_{\mu }^\nu,
\label{feqs-10} \\
	\imath \gamma^\mu \psi_{;\mu} - \mu  \psi - \frac{\partial F}{\partial\bar\psi}&=& 0,
\label{feqs-20}\\
	\imath \bar\psi_{;\mu} \gamma^\mu + \mu  \bar\psi +
	\frac{\partial F}{\partial\psi}&=& 0,
\label{feqs-21}\\
\frac{1}{\sqrt{-g}}\frac{\partial}{\partial x^\mu}\left(\sqrt{-g}g^{\mu\nu}\frac{\partial\phi}{\partial x^\nu}\right)&=&0.
\label{feqs-22}
\end{eqnarray}
The equation~\eqref{feqs-10} involves the energy-momentum tensor
 $T_{\mu}^\nu$, which can be written in a symmetric form as
\begin{align}
\label{EM}
\begin{split}
	T_{\mu}^\nu =&\frac{\imath }{4}g^{\nu\rho}\left[\bar\psi \gamma_{\mu} \psi_{;\rho}+\bar\psi\gamma_\rho\psi_{;\mu}
-\bar\psi_{;\mu}\gamma_{\rho }\psi-\bar\psi_{;\rho}\gamma_\mu\psi
\right]-\delta_\mu^\nu L_{\text{sp}}
\\
&-\partial^\nu\phi\partial_\mu\phi+\frac{1}{2}\delta_{\mu}^\nu \partial^\sigma\phi\,\partial_\sigma\phi.
\end{split}
\end{align}
Next, on account of the Dirac equations \eqref{feqs-20} and \eqref{feqs-21}, the Lagrangian \eqref{lagr_sp} takes the form:
$$
	L_{\text{sp}} = - F(S) + \frac{1}{2} \left(
		\bar\psi\frac{\partial F}{\partial\bar\psi} +
		\frac{\partial F}{\partial\psi}\psi
	\right).
$$
For our purpose,  the nonlinear term appearing here can be chosen in a simple power-law form,
$F(S) = - k(k+1)^{-1}\lambda\left(\bar\psi\psi\right)^{k+1},
$
where $k$ and $\lambda$ are some free parameters. In what follows we set $k=1$ to give
$$
	F(S) = - \frac{\lambda}{2} \left(\bar\psi\psi\right)^2.
$$
It has been demonstrated in Ref.~\cite{Dzhunushaliev:2018jhj} that the case of $\bar\lambda > 0$ corresponds to the attraction,
while the case of $\bar\lambda < 0$ -- to the repulsion.
In the absence of gravitation, classical spinor fields with such a nonlinearity have been studied, for example,
in Refs.~\cite{Finkelstein:1951zz,Finkelstein:1956,Soler:1970xp}, where it has been shown that there are regular finite energy solutions.
In turn, solitonic solutions to the nonlinear Dirac equation in a curved background have been examined in Ref.~\cite{Mielke:2017nwt}
(see also references therein).

Here we consider only spherically symmetric configurations, for which one can choose the spacetime metric in the form 
\begin{equation}
	ds^2 = e^{A(r)} dt^2 - B(r) e^{-A(r)}\left[dr^2 + \left(r^2+r_0^2\right) \left(d \theta^2 + \sin^2 \theta d \varphi^2\right)\right],
\label{metric}
\end{equation}
where $e^{A(r)}=1-2 G m(r)/r$, and the function $m(r)$ corresponds to the current mass of the configuration
enclosed by a sphere with circumferential radius $r$. The parameter $r_0$ characterises the throat; without the spinor field, it is the radius of the throat
 (see below).

For a description of the spinor field, we take the following stationary {\it Ansatz} 
compatible with the spherically symmetric line element \eqref{metric} (see, e.g., Refs.~\cite{Soler:1970xp,Li:1982gf,Li:1985gf,Herdeiro:2017fhv}):
\begin{equation}
	\psi^T =\frac{e^{-\imath \Omega t}}{\sqrt{2}}\,  \begin{Bmatrix}
		\begin{pmatrix}
			0 \\ - u \\
		\end{pmatrix},
		\begin{pmatrix}
			u \\ 0 \\
		\end{pmatrix},
		\begin{pmatrix}
			\imath v \sin \theta e^{- \imath \varphi} \\ - \imath v \cos \theta \\
		\end{pmatrix},
		\begin{pmatrix}
			- \imath v \cos \theta \\ - \imath v \sin \theta e^{\imath \varphi} \\
		\end{pmatrix}
	\end{Bmatrix},
\label{spinor}
\end{equation}
where $\Omega$ is the spinor frequency and
$u(r)$ and $v(r)$ are two real functions.
This  {\it Ansatz} ensures that the spacetime remains static. 
Each row of this  {\it Ansatz} describes a  spin-$\frac{1}{2}$ fermion, and these two fermions have the same masses $\mu$ and opposite spins.
Although the energy-momentum tensors of these fermions are not spherically symmetric, their sum ensures a spherically symmetric energy-momentum tensor.

Converting the {\it Ansatz}~\eqref{spinor} into spherical coordinates (see Ref.~\cite{Dzhunushaliev:2018jhj}) and  
substituting the resulting expression and the metric  \eqref{metric} into the field equations
\eqref{feqs-10}, \eqref{feqs-20}, and  \eqref{feqs-22}, one can obtain the following set of equations:
\begin{align}
& A^{\prime\prime}+\left(\frac{2x}{x^2+x_0^2}+\frac{1}{2}\frac{B^\prime}{B}\right)A^\prime+2 e^{-3 A/2}B
\left[e^{A/2}\left(\bar{u}^2-\bar{v}^2\right)-2\bar\Omega\left(\bar{u}^2+\bar{v}^2\right)\right]=0,
\label{Eq_A}\\
& B^{\prime\prime}+\left(\frac{3 x}{x^2+x_0^2}-\frac{1}{2}\frac{B^\prime}{B}\right)B^\prime+4 e^{-A/2}B^{3/2}\frac{\bar u \bar v}{\sqrt{x^2+x_0^2}}+
4 e^{-3 A/2}B^2\left[e^{A/2}\left(\bar{u}^2-\bar{v}^2\right)-\bar\Omega\left(\bar{u}^2+\bar{v}^2\right)\right]=0,
\label{Eq_B}\\
& \bar u^\prime+\left(\frac{x}{x^2+x_0^2}-\frac{1}{\sqrt{x^2+x_0^2}}-\frac{1}{4}A^\prime+\frac{1}{2}\frac{B^\prime}{B}\right)\bar u+
e^{-A}\sqrt{B}\Big\{\bar\Omega+e^{A/2}\left[1-\bar\lambda\left(\bar{u}^2-\bar{v}^2\right)\right]\Big\}\bar{v}=0,
\label{Eq_u}\\
& \bar v^\prime+\left(\frac{x}{x^2+x_0^2}+\frac{1}{\sqrt{x^2+x_0^2}}-\frac{1}{4}A^\prime+\frac{1}{2}\frac{B^\prime}{B}\right)\bar v-
e^{-A}\sqrt{B}\Big\{\bar\Omega-e^{A/2}\left[1-\bar\lambda\left(\bar{u}^2-\bar{v}^2\right)\right]\Big\}\bar{u}=0,
\label{Eq_v}\\
& \left[\sqrt{B}\left(x^2+x_0^2\right)\bar\phi^\prime\right]^\prime=0,
\label{Eq_phi}
\end{align}
where the prime denotes differentiation with respect to the radial coordinate.
Here,  Eq.~\eqref{Eq_A} is obtained from the Einstein equations~\eqref{feqs-10} as the combination 
 $\left[\left(_t^t\right)-\left(_r^r\right)\right]$, where the second derivative $B^{\prime\prime}$ is eliminated using
 the~$\left(_\theta^\theta\right)$-component. In turn,  Eq.~\eqref{Eq_B} is the combination  $\left[\left(_r^r\right)+\left(_\theta^\theta\right)\right]$.
The above equations are written in terms of the following dimensionless variables and parameters:
\begin{equation}
\label{dmls_var}
	x =  \mu r, \quad
	\bar \Omega = \frac{\Omega}{\mu}, \quad
	\left(\bar u, \bar v\right) = \sqrt{\frac{4\pi G}{\mu}}\left( u, v\right),\quad
	  \bar \phi = \sqrt{8\pi G}\phi,\quad
	\bar \lambda =  \frac{\lambda}{4\pi G} 	 .  
\end{equation}

Note that Eqs.~\eqref{Eq_A} and \eqref{Eq_B} do not explicitly contain the scalar field, since it was eliminated  by combining the corresponding components of the Einstein equations.
In turn, the integration of the equation for the scalar field ~\eqref{Eq_phi} yields
\begin{equation}
\label{phi_first_int}
\bar\phi^\prime=\frac{\sqrt{\bar D}}{\sqrt{B}\left(x^2+x_0^2\right)} ,
\end{equation}
where the dimensionless integration constant $\bar D\equiv 8\pi G\mu^2 D$ represents the scalar charge of the ghost field. By substituting this expression into the~$\left(^r_r\right)$-component of the Einstein equations~\eqref{feqs-10},
one can find 
\begin{align}
\bar D&=-\frac{1}{2}\left(x^2+x_0^2\right)^2 \frac{B^{\prime 2}}{B}-2 x \left(x^2+x_0^2\right)B^\prime+\frac{1}{2}\left[4 x_0^2+\left(x^2+x_0^2\right)^2 A^{\prime 2}\right]B\nonumber\\
&+2 e^{-3 A/2}\left(x^2+x_0^2\right)^2 B^2\Big\{2\bar\Omega\left(\bar u^2+\bar v^2\right)+e^{A/2}\left(\bar u^2-\bar v^2\right)\left[\bar\lambda\left(\bar u^2-\bar v^2\right)-2\right]
\Big\}-8 e^{-A/2}\left(x^2+x_0^2\right)^{3/2}B^{3/2}\bar u\bar v .
\label{constr}
\end{align}
Below we employ the condition $\bar D=\text{const}$ to monitor the quality of the numerical solutions. The variation of the constant $\bar D$ as computed from
Eq.~\eqref{constr}  is typically less than $10^{-4}$.

\section{Numerical solutions}
\label{num_sol}

In this section, we discuss the numerical solutions of the equations~\eqref{Eq_A}-\eqref{Eq_phi} and the
physical properties of configurations described by these solutions.

\subsection{Asymptotic behavior and boundary conditions}

We seek globally regular finite-energy solutions of the set of five ordinary differential equations~\eqref{Eq_A}-\eqref{Eq_phi}.
Even before obtaining numerical solutions, it is possible to estimate their asymptotic behavior, keeping in mind that
for the functions $\bar u$ and $\bar v$ we will seek solutions that decay exponentially with distance as $x\to \pm \infty$. 
It is seen from the form of Eqs.~\eqref{Eq_u} and \eqref{Eq_v} that, due to the presence of the term $\left(x^2+x_0^2\right)^{-1/2}$, they are not $Z_2$-symmetric. 
Consistent with this, we will seek solutions that are asymmetric with respect to the point $x=0$.
Then the corresponding asymptotic behavior of the spinor fields is of the form
$$
\bar{u}\approx \bar{u}_{\pm\infty}\frac{e^{\mp\sqrt{1-\bar{\Omega}^2}x}}{x}+\cdots, 
\quad \bar{v}\approx \bar{v}_{\pm\infty}\frac{e^{\mp\sqrt{1-\bar{\Omega}^2}x}}{x}+\cdots,
$$
where $\bar{u}_{\pm\infty}, \bar{v}_{\pm\infty}$ are integration constants for $x\to \pm \infty$, respectively. 
In turn, for the metric functions $A$ and $B$, one can find from Eqs.~\eqref{Eq_A} and~\eqref{Eq_B} as  $x\to \pm\infty$
\begin{equation}
 A\approx \mp\frac{2 \bar M_\pm}{x}+\cdots, \quad B \to 1+\cdots .
\label{A_asympt}
\end{equation}
Here $\bar M_+$ corresponds to a total mass of the systems under consideration
as measured by an observer when $x\to +\infty$ and $\bar{M}_-$ is the mass as measured when $x\to -\infty$.

The asymptotic behavior of the scalar field follows from Eq.~\eqref{phi_first_int} in the form
$$
\bar{\phi}\approx \bar{\phi}_{\pm \infty}-\frac{\sqrt{\bar D}}{x} +\cdots ,
$$
where $\bar{\phi}_{\pm \infty}$ are two integration constants corresponding to the values of the field as $x\to \pm \infty$, respectively.

Consistent with the above asymptotic behavior, the corresponding boundary conditions can be chosen in the form
\begin{equation}
\label{BCs}
A(x\to \pm \infty)=0, \quad B(x\to \pm \infty)=1, \quad \bar u(x\to \pm \infty)=0, \quad \bar v(x\to \pm \infty)=0, \quad \bar \phi(x\to \pm \infty)=\bar{\phi}_{\pm \infty}.
\end{equation}
These boundary conditions imply that the spacetime is asymptotically flat. 

\subsection{Numerical method}

We solve the set of mixed order differential equations \eqref{Eq_A}-\eqref{Eq_phi}  
subject to the boundary conditions~\eqref{BCs} and use the constraint equation \eqref{constr} to verify the accuracy of calculations.

In order to map the infinite range of the radial variable $x$ 
to the finite interval, we introduce the compactified radial coordinate $\bar x$,
\begin{equation}
    \bar x = \frac{2}{\pi}\arctan{\left(\frac{x}{c_k}\right)} \, ,
\label{comp_coord}
\end{equation}
which maps the infinite region $(-\infty;\infty)$ onto the finite interval $[-1; 1]$. Here $c_k$ is an arbitrary constant which is used to adjust the contraction of
the grid. In our calculations, we typically take $c_k\in [0.1,3]$.

Technically, Eqs.~\eqref{Eq_A}-\eqref{Eq_phi} are discretized on a grid
consisting usually of about 1000 grid points, but in some cases even 3000 or more grid points have been used.
The emerging system of nonlinear algebraic equations 
is then solved using a modified Newton method.
The underlying linear system is solved 
with the Intel MKL PARDISO sparse direct solver \cite{pardiso} 
and the CESDSOL library\footnote{Complex Equations-Simple Domain 
partial differential equations SOLver, a C++ package developed by I.~Perapechka,
see Refs.~\cite{Herdeiro:2019mbz,Herdeiro:2021jgc}.}.
The package provides an iterative procedure to obtain an exact solution starting from some initial guess configuration. 

As such a configuration, one can take, for example, a system described by a solution of the Dirac equations in the background of a pure wormhole with zero mass.
Solutions describing such a wormhole can be obtained by solving Eqs.~\eqref{Eq_A} and \eqref{Eq_phi} in the absence of the spinor fields
 ($\bar{u}=\bar{v}=0$) and with $B=1$ in the form~\cite{Bronnikov:1973fh,Ellis:1973yv,Ellis:1979bh,Gonzalez:2008wd}
$$
\bar\phi=\sqrt{2\left(1+\gamma_1^2\right)}\arctan{\left(\frac{x}{x_0}\right)}+\bar\phi_0, \quad
A=2\gamma_1\arctan{\left(\frac{x}{x_0}\right)}+2\gamma_0 ,
$$
where $\gamma_0, \gamma_1,$ and $\bar\phi_0$ are some constants. Note that the constant $ \bar\phi_0$ has no physical meaning since only the gradient
of the scalar field enters the equations. In turn, the constants  $\gamma_0$ and $\gamma_1$ determine the mass of such wormhole.

Here we will consider the simplest case of a zero mass wormhole $\bar{M}_+=\bar{M}_-=0$ for which $\gamma_0=\gamma_1=0$~\cite{Gonzalez:2008wd}.
Then, by solving the Dirac equations~\eqref{Eq_u} and \eqref{Eq_v} in the background of the metric $A=0$ and $B=1$, 
one can use the configuration obtained as the aforementioned initial guess configuration.

\subsection{Energy conditions}

 The violation of the null  and weak energy conditions, needed for ensuring nontrivial topology in the system, 
 implies the violation of the following inequalities,
$$
    T_{\mu\nu} k^\mu k^\nu \geq 0 \quad \text{and}\quad T_{\mu\nu} V^\mu V^\nu \geq 0 ,
$$
for any null vector $k^\mu$, $g_{\mu\nu}k^\mu k^\nu =0$, 
and for any timelike vector $V^\mu$, $g_{\mu\nu}V^\mu V^\nu >0$, 
respectively (for a review, see, e.g., Ref.~\cite{Visser}).
The weak energy condition also implies $T_0^0\geq 0$.

Since the violation of the null energy condition implies the violation of the weak and the strong energy conditions, here
we address only the null energy condition. The latter can be reexpressed by making use of the Einstein equations~\eqref{feqs-10} in the form
$$
G_t^t-G_r^r \geq 0 \quad \text{and} \quad G_t^t-G_\theta^\theta \geq 0 .
$$
The null energy condition is violated when one or both of these conditions do not hold in some region of the spacetime.
Note that for the configurations considered below only the first of these conditions is always violated.

\subsection{Mass and the circumferential radius}

We consider Dirac-star-plus-wormhole configurations that are asymptotically flat and
asymmetric with respect to the center $x=0$. The key point here is the behavior 
of the circumferential radius $\bar R(x)$ defined as
\begin{equation}
\bar R^2\equiv g_{\theta\theta}=B e^{-A}\left(x^2+x_0^2\right) .
\label{circ_radius}
\end{equation}
Asymptotic flatness requires that $\bar R(x) \to |x|$ for large $|x|$.
Because of the  asymmetry of the systems, 
the center of the configurations located at $x=0$ should not in general already correspond to an extremum
of $\bar R(x)$. Depending on the specific values of the system parameters, the  extremum
of $\bar R(x)$ can be located both to the left and to the right of the point $x=0$. 
If $\bar R(x)$ has only one global minimum at  some point $x=x_{\text{extr}}$, then $x_{\text{extr}}$ corresponds to the throat
of the wormhole $\bar R_{\text{th}}=\text{min}\{\bar R(x)\}$ (a single-throat system).
If, on the other hand, $\bar R(x)$ has a local maximum at  $x=x_{\text{extr}}$,
then this point corresponds to an equator $\bar R_{\text{eq}}=\text{max}\{\bar R(x)\}$.
This then implies that there are (at least) two minima of $\bar R(x)$ (on account of the asymmetry of the system,
one of them will be global and another one~-- local), located, in general, asymmetrically to the left and to the right of the maximum. 
In the case of two such minima, the wormhole will have a double throat
surrounding a belly (see, e.g., Refs.~\cite{Charalampidis:2013ixa,Hauser:2013jea,Dzhunushaliev:2014mza}).

Let us now turn to the total mass of the configurations in question. Since the objects with nontrivial topology under consideration are asymmetric with respect to the center $x=0$,
a distant observer located at $x\to \pm \infty$ measures different values of the mass $\bar{M}_+$ and $\bar{M}_-$ at infinity, see Eq.~\eqref{A_asympt}.
They correspond  to a dimensionless ADM mass of the configuration $\bar M$ given by
\begin{equation}
\label{expres_mass}
\bar M_\pm\equiv \mu M_\pm/M_p^2=\pm\frac{1}{2}\lim_{x\to\pm\infty}x^2\partial_x e^A = \pm\frac{c_k}{\pi}\lim_{\bar x\to \pm 1}\partial_{\bar x} A ,
\end{equation}
where $M_p$ is the Planck mass and the last expression in the above formula gives the mass in terms of the
compactified coordinate $\bar x$ from Eq.~\eqref{comp_coord}.

Alternatively, the mass of the system can also be found from the $(^t_t)$-component
of the energy-momentum tensor~\eqref{EM},
\begin{equation}
m(r)=\frac{1}{2G} R_{\text{extr}}+4\pi\int_{R_{\text{extr}}}^r T_t^t R^2 dR .
\label{m_current}
\end{equation}
Here the  circumferential radius $R_{\text{extr}}$ corresponds either to 
the radius of the wormhole throat  $R_{\text{th}}$ (for single-throat systems) or to the radius of the equator $R_{\text{eq}}$
(for double-throat systems).
In terms of the dimensionless variables~\eqref{dmls_var} the expression \eqref{m_current} can be recast in the form
\begin{equation}
\bar m(x)\equiv G \mu m(r)=\frac{\bar{R}_{\text{extr}}}{2}+\frac{1}{4}\int_{x_{\text{extr}}}^{x}\bar T_t^t \bar R^2 \frac{d\bar R}{d x^\prime} d x^\prime ,
\label{m_current_dmls}
\end{equation}
where $x_{\text{extr}}$ is the point on the $x$-axis where a throat or an equator are located.
The expression~\eqref{m_current_dmls} can be employed to monitor the accuracy of the numerical calculations.

\subsection{Results of numerical calculations}

\begin{figure}[t]
    \begin{center}
        \includegraphics[width=.49\linewidth]{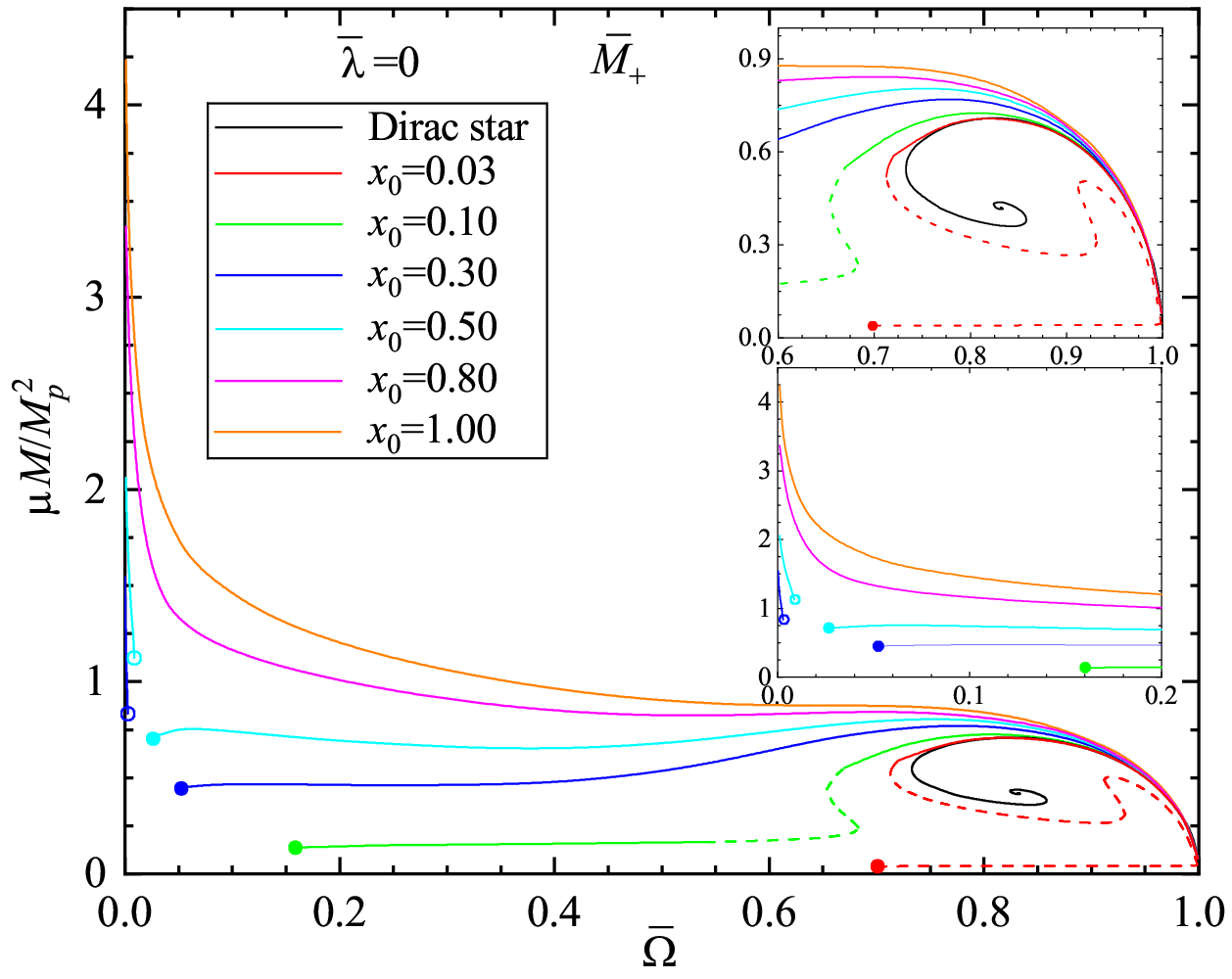}
        \includegraphics[width=.49\linewidth]{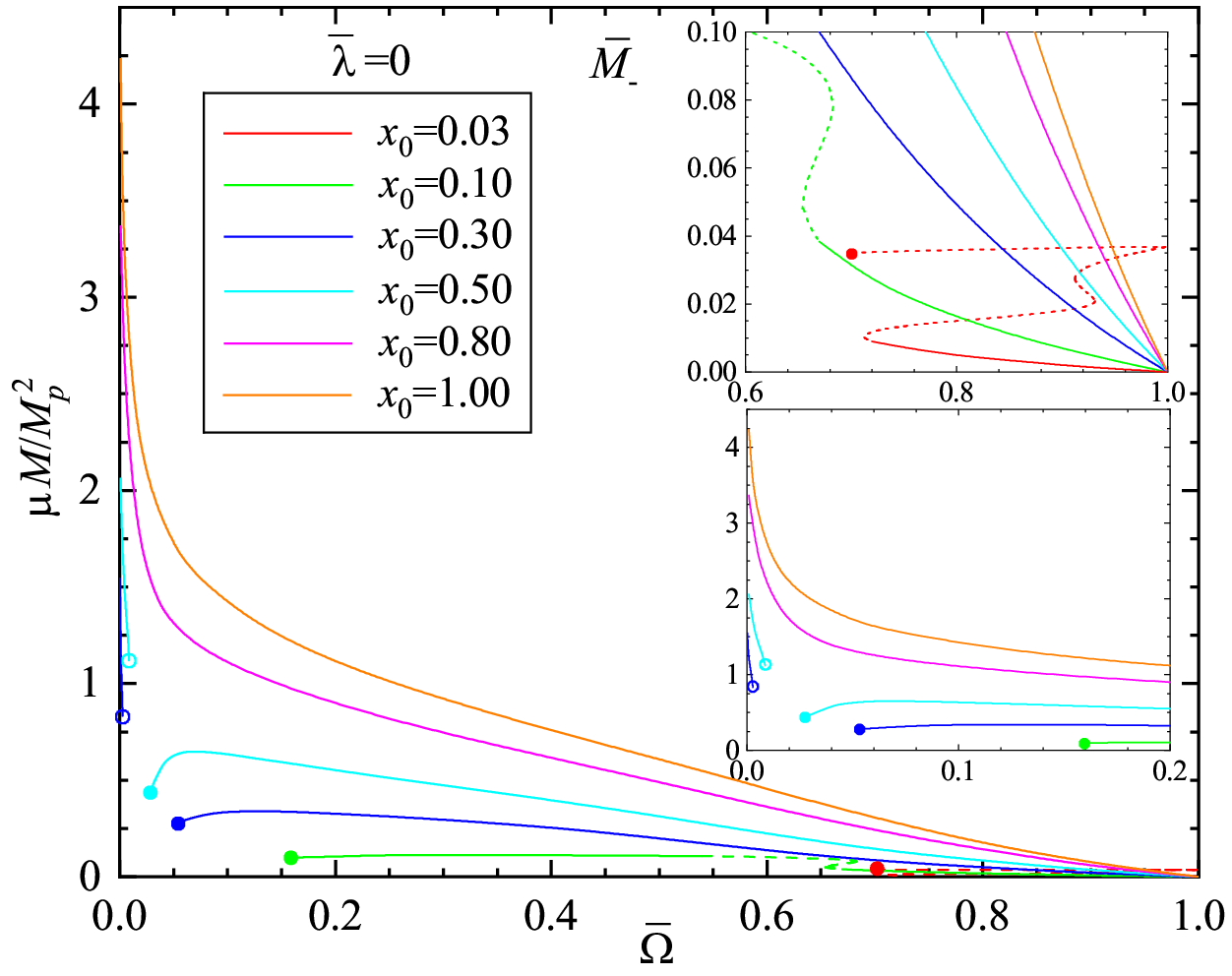}
    \end{center}
    \vspace{-.5cm}
    \caption{The dimensionless Dirac-star-plus-wormhole total mass $\bar M_\pm$ as a function of
the parameter $\bar\Omega$ for a linear spinor field ($\bar{\lambda}=0$) and different values of $x_0$. 
The circles mark the configurations with the limiting values $\bar{\Omega}_1$ (solid circles) and $\bar{\Omega}_0$ (open circles).
The insets show the behavior of the curves for large and small $\bar{\Omega}$. The dashed lines mark the parts of the curves for the mixed configurations where the binding energy is negative
(energetically unstable systems).
    }
    \label{fig_Mass_Omega_Lambda_0}
\end{figure}

\begin{figure}[t]
    \begin{center}
        \includegraphics[width=.49\linewidth]{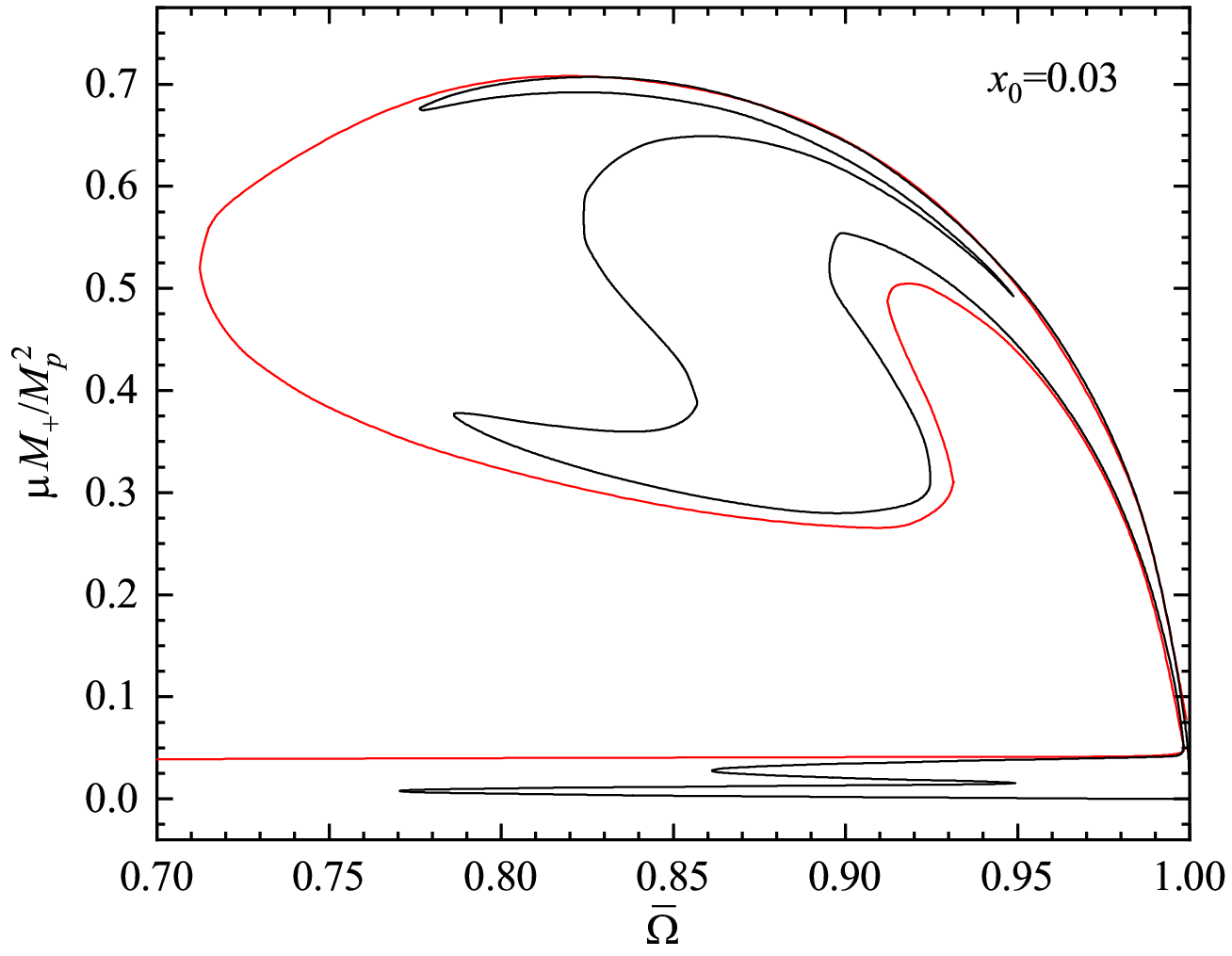}
        \includegraphics[width=.49\linewidth]{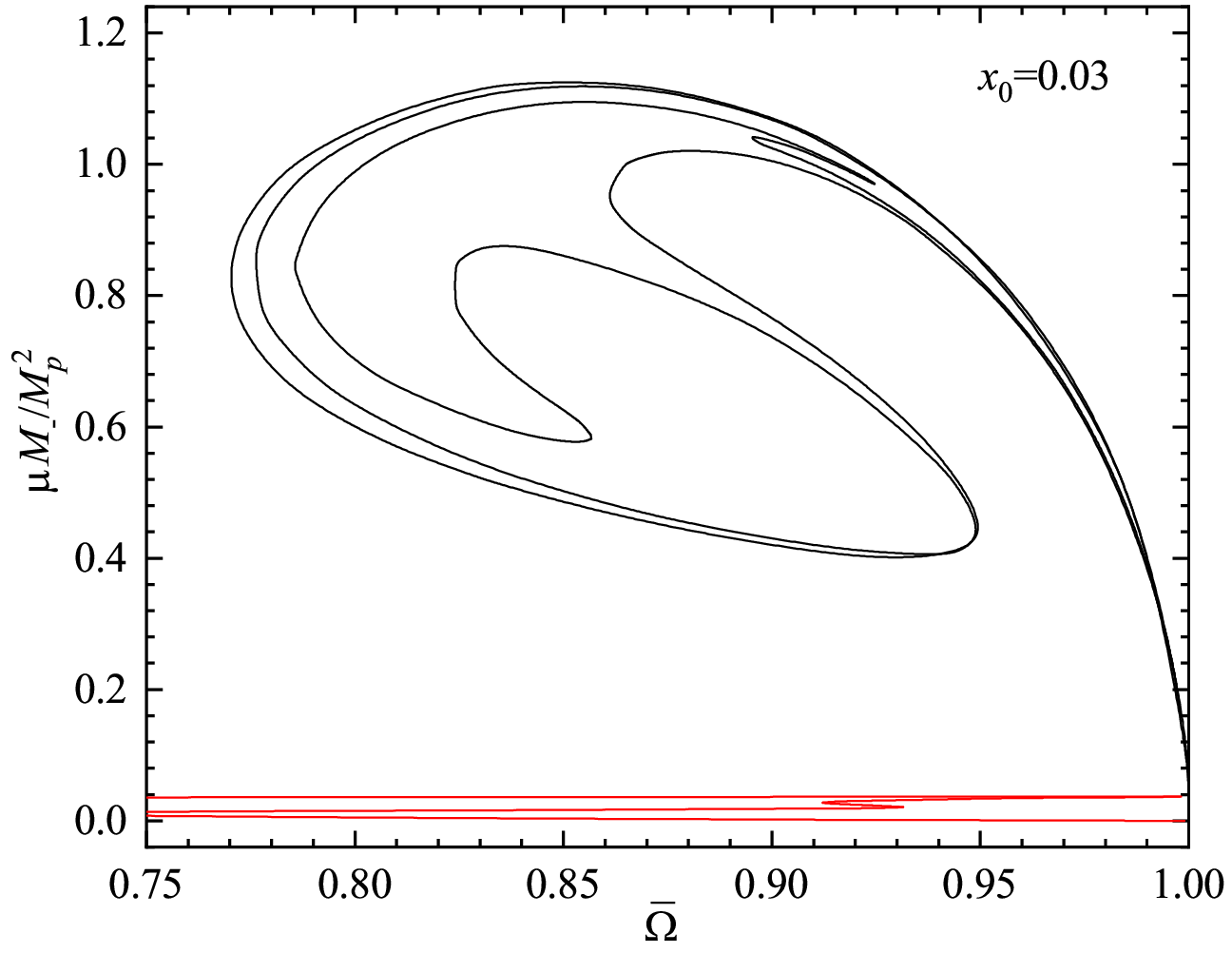}
    \end{center}
    \vspace{-.5cm}
    \caption{The dimensionless Dirac-star-plus-wormhole total mass $\bar M_\pm$ as a function of
the parameter $\bar\Omega$ for a linear spinor field
 for the family of solutions of Ref.~\cite{Hao:2023igi} (solid black lines) and
for the present family of solutions (red solid lines).
    }
    \label{fig_Mass_Omega_Lambda_0_comp}
\end{figure}

\begin{figure}[!]
    \begin{center}
        \includegraphics[width=1.\linewidth]{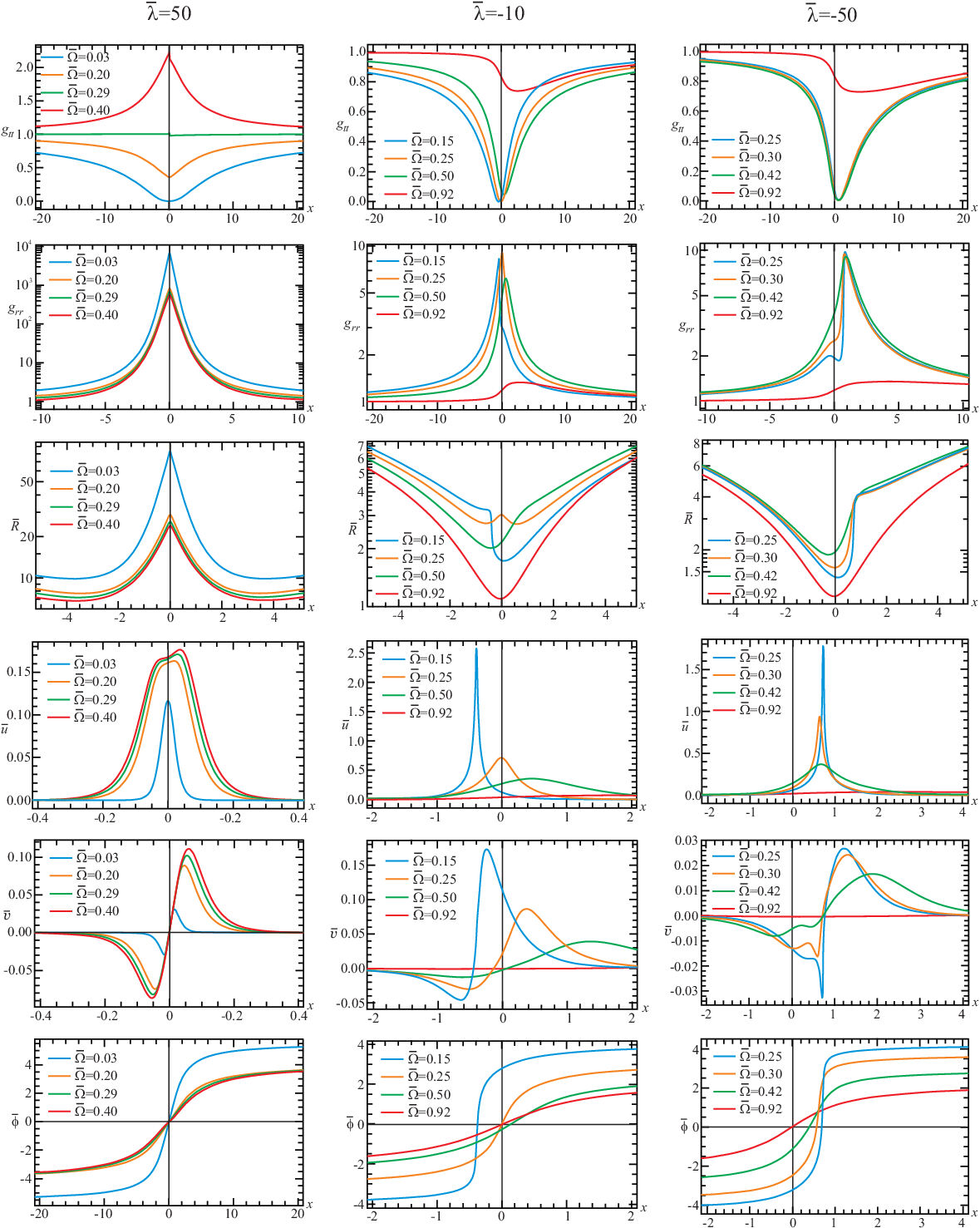}
    \end{center}
    \vspace{-.5cm}
    \caption{Typical solutions for different values of the nonlinearity parameter $\bar{\lambda}$ for a fixed $x_0=1$.
       }
    \label{fig_plots_sols}
\end{figure}

The boundary conditions in the form~\eqref{BCs} imply solving a two-point boundary value problem for the set of equations~\eqref{Eq_A}-\eqref{Eq_phi}.
This set is solved by assigning different values of the free parameters of the system $x_0$, $\bar\lambda$, and $\bar\Omega$ to obtain regular solutions.

\subsubsection{The case of a linear spinor field}
\label{linear_field}

To begin with, let us consider the case of a linear Dirac field when $\bar\lambda=0$. 
This case was examined earlier in Ref.~\cite{Hao:2023igi}. Here we  find a new family of solutions,
which is characterized by  a qualitatively different behavior of dependencies of the total mass on the spinor frequency. 
Namely, Fig.~\ref{fig_Mass_Omega_Lambda_0} shows the dependencies of the masses $\bar{M}_+$ and $\bar{M}_-$
on the frequency $\bar{\Omega}$ for different values of the throat parameter $x_0$. We start with the case of a Dirac star with trivial topology, 
for which the mass curve $\bar{M}_+$ has a characteristic spiral-like form~\cite{Herdeiro:2017fhv}. Upon adding the ghost scalar field,
as $x_0$ increases, the spiral unwinds, and eventually degenerates to a monotonically increasing function.
All the solutions start at $\bar\Omega=1$ from the limiting Ellis wormhole solution with zero mass. However, as seen from the figure, depending on the value of $x_0$,
two types of the behavior of solutions are possible.

The configurations of the first type, for which $x_0=0.03, 0.1, 0.3, 0.5$. The systems of such type,
if one starts the solution from the frequency $\bar{\Omega}\to 1$ and gradually decreases it,
always possess some limiting value $\bar{\Omega}_{1}$, for which one can still perform numerical calculations.
On the other hand, if one starts the solution from the frequency $\bar{\Omega}\to 0$ and gradually increases it,
there also exists some other limiting value  $\bar{\Omega}_{0}<\bar{\Omega}_{1}$,  for which one can still perform computations. 
This is illustrated in the insets of Fig.~\ref{fig_Mass_Omega_Lambda_0} for the values $x_0=0.5$ and $x_0=0.3$, where one can see a gap in $\bar{\Omega}$
equal to the difference  $\left(\bar{\Omega}_{1}-\bar{\Omega}_{0}\right)$ 
(for the values  $x_0=0.1$ and $x_0=0.03$ the corresponding curves do exist only at extremely small values of
 $\bar{\Omega} \to 0$, and they cannot be shown in the figure).

It is important to note that when the frequency tends to both aforementioned limiting values $\bar{\Omega}_1$ and $\bar{\Omega}_0$, as well as when $\bar{\Omega}\to 0$,  
the metric function $g_{tt}\equiv e^A \to 0$. The latter assumes the possible presence of a horizon for such systems. Depending on the limiting value to which $\bar{\Omega}$ tends,
the dimensionless Kretschmann scalar $\bar{K}\equiv K/\mu^4=\bar R_{\alpha\beta\mu\nu}\bar R^{\alpha\beta\mu\nu}$ behaves as follows: 
when $\bar{\Omega}\to 0$, we always have $\bar{K}\to 0$; when $\bar{\Omega}\to \bar{\Omega}_{0}$, the Kretschmann scalar possesses finite values;
when $\bar{\Omega}\to \bar{\Omega}_{1}$, the Kretschmann scalar diverges, i.e., such systems possess a singularity~-- a possible singular horizon 
(for a detailed consideration of the behavior of solutions in this limit, see the case of a nonlinear spinor field considered below). 
Note that in all these cases we deal with double-throat systems.

As  $x_0$ increases and crosses some threshold value, there arise systems of the second type, which are represented in Fig.~\ref{fig_Mass_Omega_Lambda_0} 
by the families of solutions with $x_0=0.8$ and $x_0=1$. Such systems are characterized by the absence of a gap in the frequency: as $\bar{\Omega}$ increases, the mass decreases monotonically. 
In turn, the Kretschmann scalar, tending to 0 as  $\bar{\Omega} \to 0$ (where $g_{tt}\equiv e^A \to 0$ as well), also increases monotonically up to some finite value at
 $\bar{\Omega} \to 1$ (where  $g_{tt}\to 1$ and we deal with the limiting Ellis wormhole solution).

Note that for the systems of both types, it is characteristic that  (i)~As  $\bar{\Omega}\to 0$, the configurations always have two throats separated by an equator, whose radius 
 $\bar R_{\text{eq}}$ increases very rapidly. One may expect that it will eventually diverge; this corresponds to the fact that a possible horizon with infinite area is located at the equator.
 Such configurations are referred to as ``cold black holes''~\cite{Bronnikov:2006qj}.
(ii)~For large values of $\bar{\Omega}$, the system possesses only one throat with the circumferential radius $\bar R_{\text{th}}$, 
and the throat is located to the left of the center $x=0$. However, depending on the value of $x_0$, there is some value of $\bar{\Omega}$ 
where the single-throat configurations become systems with two throats separated by an equator with the circumferential radius $\bar R_{\text{eq}}$. 
In general, these throats are located asymmetrically with respect to the center  $x=0$ and they have unequal
circumferential radii $\bar R_{\text{th}+}\neq \bar R_{\text{th}-}$.

To conclude this section, let us compare mass curves for the solutions of Ref.~\cite{Hao:2023igi} and for the solutions of the present paper
for a particular case with the throat parameter $x_0=0.03$. As seen from Fig.~\ref{fig_Mass_Omega_Lambda_0_comp}, 
 the family of solutions of Ref.~\cite{Hao:2023igi} is characterized by a considerably different qualitative behavior of the mass curves. 
Here, as in the case of the family considered above, the solutions start at $\bar\Omega=1$ from the limiting Ellis wormhole solution with zero mass.
However, this family is characterized by a considerably more complicated structure of branches with large number of turning points.
Eventually, unlike the solutions of the first family, here the solutions return back to the starting point $\bar\Omega=1$. 
In this case,  the Kretschmann scalar remains always finite, i.e., the configurations under consideration are always regular,
in contrast to the systems from the above family of solutions when a fast increase of $\bar{K}$ for $\bar{\Omega}\to \bar{\Omega}_1$ occurs.

\subsubsection{The case of a nonlinear spinor field}
\label{nonlinear_field}

Let us now turn to a consideration of systems with a nonlinear spinor field. To demonstrate a characteristic behavior of the solutions and changes in physical parameters of such systems
depending on the value of the nonlinearity parameter  $\bar\lambda$, here we restrict ourselves to a consideration of only one fixed value of the throat parameter $x_0=1$. 
In this case, varying $\bar\Omega$ for each fixed value of $\bar\lambda$,
we obtain families of Dirac star solutions harboring a wormhole at their core. We will demonstrate that, depending on the magnitude of the parameter  $\bar\lambda$,
such families of solutions start at $\bar\Omega=1$ either from the limiting Ellis wormhole solution or not. 
In turn,  as in the case of a linear spinor field, as  $\bar\Omega$ decreases, the solutions either can exist for all $\bar\Omega$ or there is a gap in the frequency caused by the presence of limiting values  
 $\bar{\Omega}_1$ and $\bar{\Omega}_0$.

\begin{figure}[t]
    \begin{center}
        \includegraphics[width=.49\linewidth]{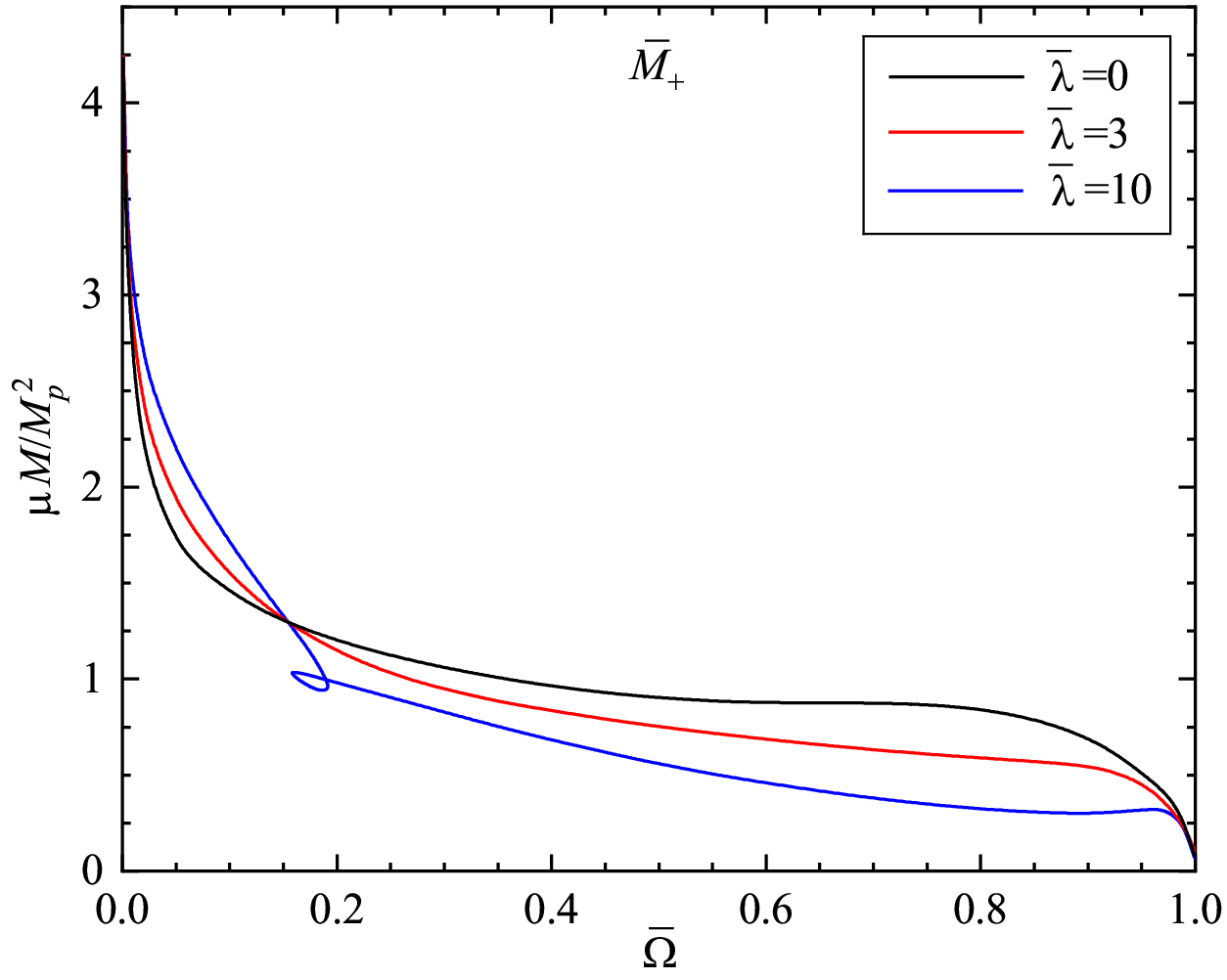}
        \includegraphics[width=.49\linewidth]{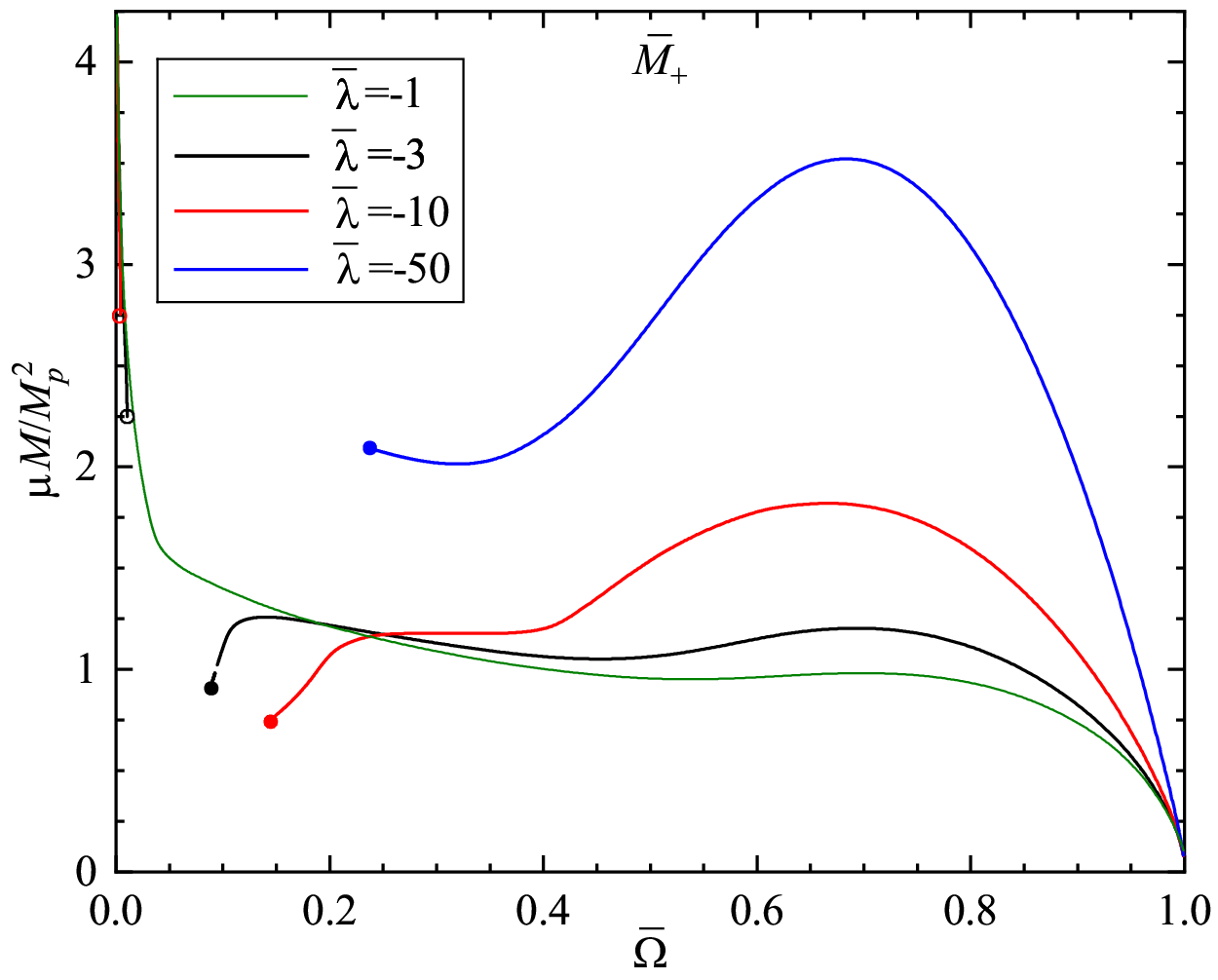}
        \includegraphics[width=.49\linewidth]{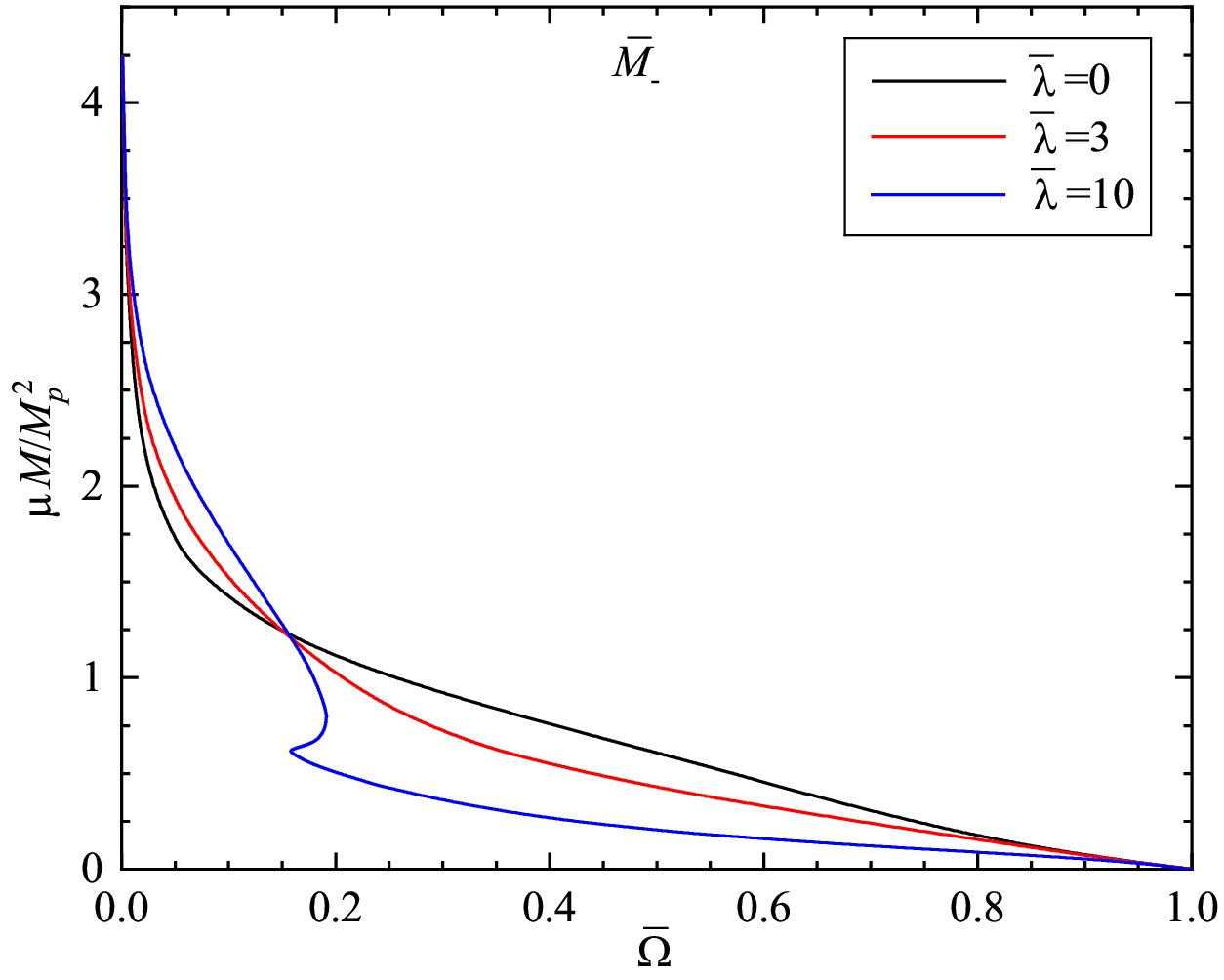}
        \includegraphics[width=.49\linewidth]{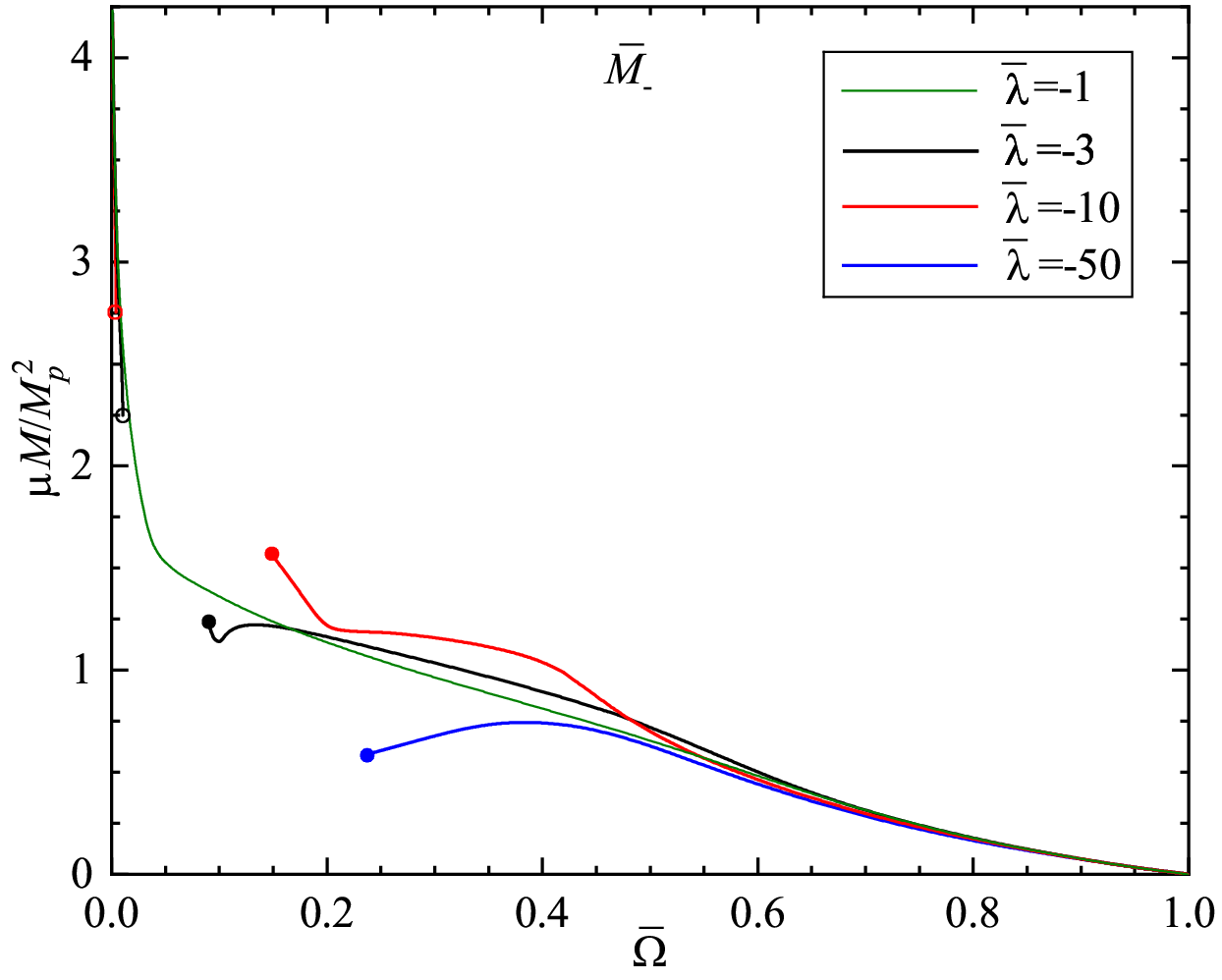}
         \end{center}
    \vspace{-.5cm}
    \caption{The dimensionless Dirac-star-plus-wormhole total mass $\bar M_\pm$ as a function of
$\bar\Omega$ for $x_0=1$ and different $\bar{\lambda}$. 
The circles in the right panels mark the configurations with the limiting values $\bar{\Omega}_1$ (solid circles) and $\bar{\Omega}_0$ (open circles).
    }
    \label{fig_Mass_Omega_Lambda}
\end{figure}

\begin{figure}[t]
    \begin{center}
        \includegraphics[width=.49\linewidth]{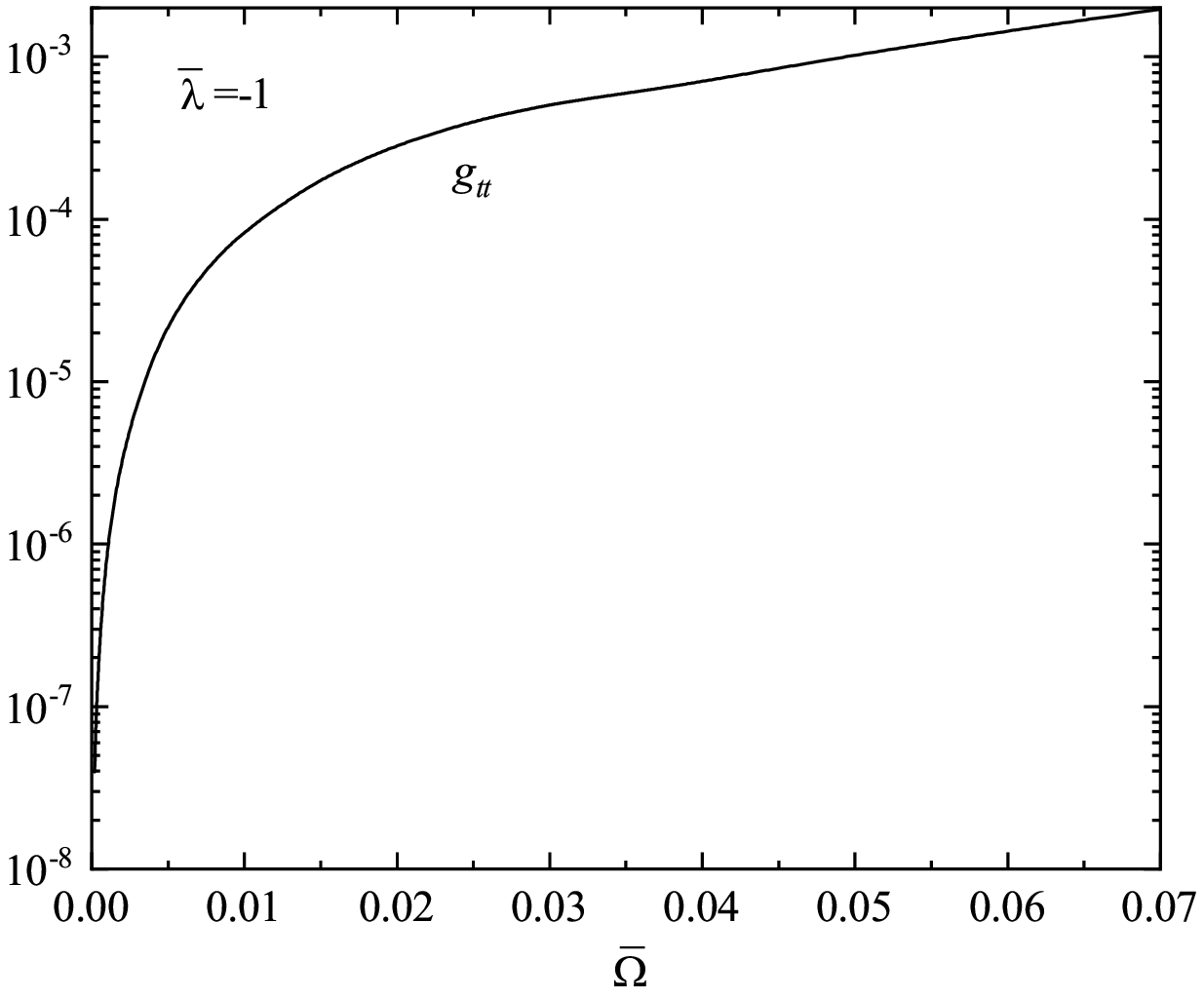}
        \includegraphics[width=.49\linewidth]{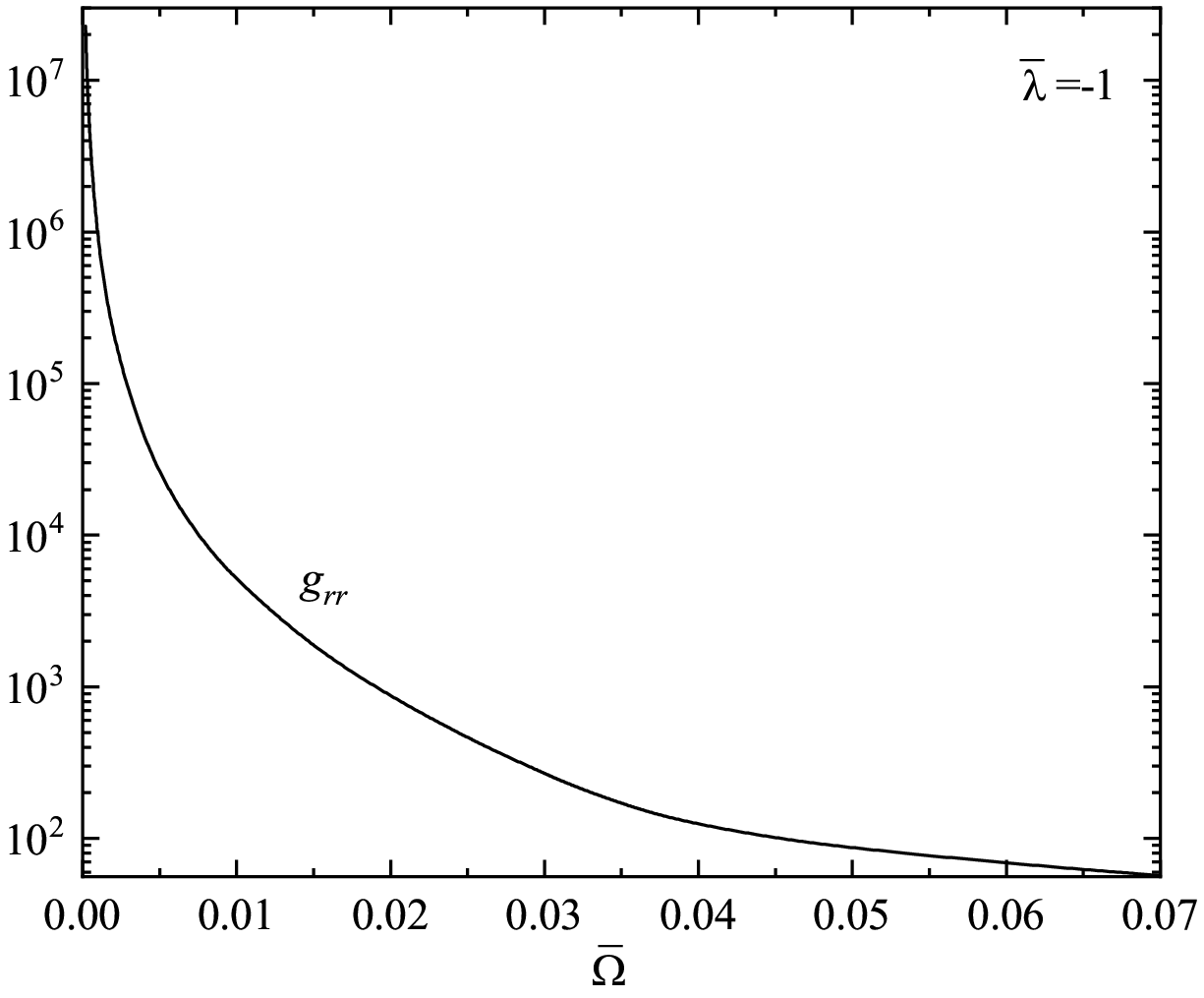}
      \end{center}
    \vspace{-.5cm}
    \caption{The characteristic behavior of the metric functions $g_{tt}\equiv e^A$ (minimum values)
    and $g_{rr}\equiv B e^{-A}$ (maximum values) for small $\bar\Omega$ for the systems with $\bar\lambda \geq \bar\lambda_{\text{tv}}$.
    }
    \label{fig_gtt_grr}
\end{figure}

\begin{figure}[t]
    \begin{center}
        \includegraphics[width=.49\linewidth]{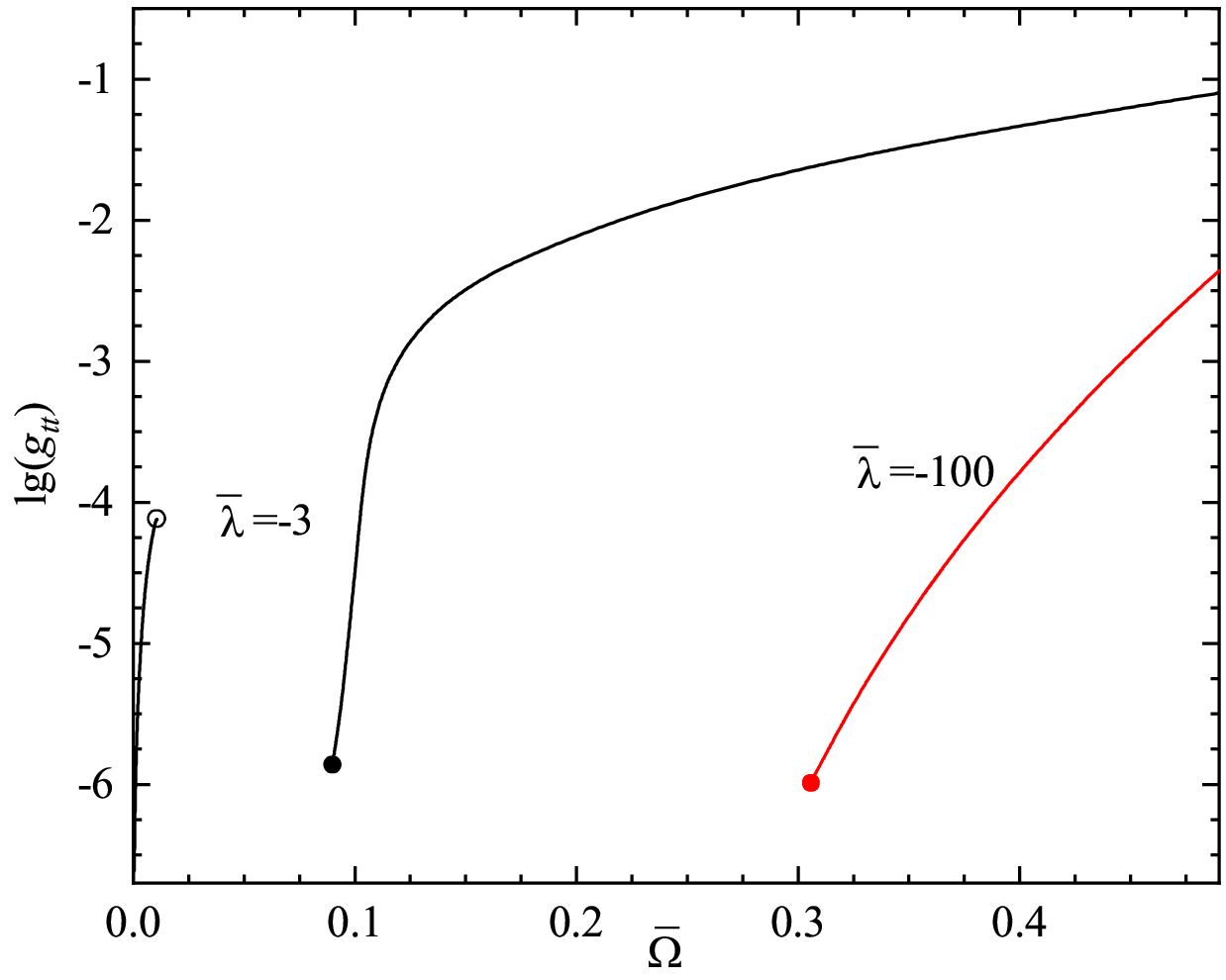}
        \includegraphics[width=.49\linewidth]{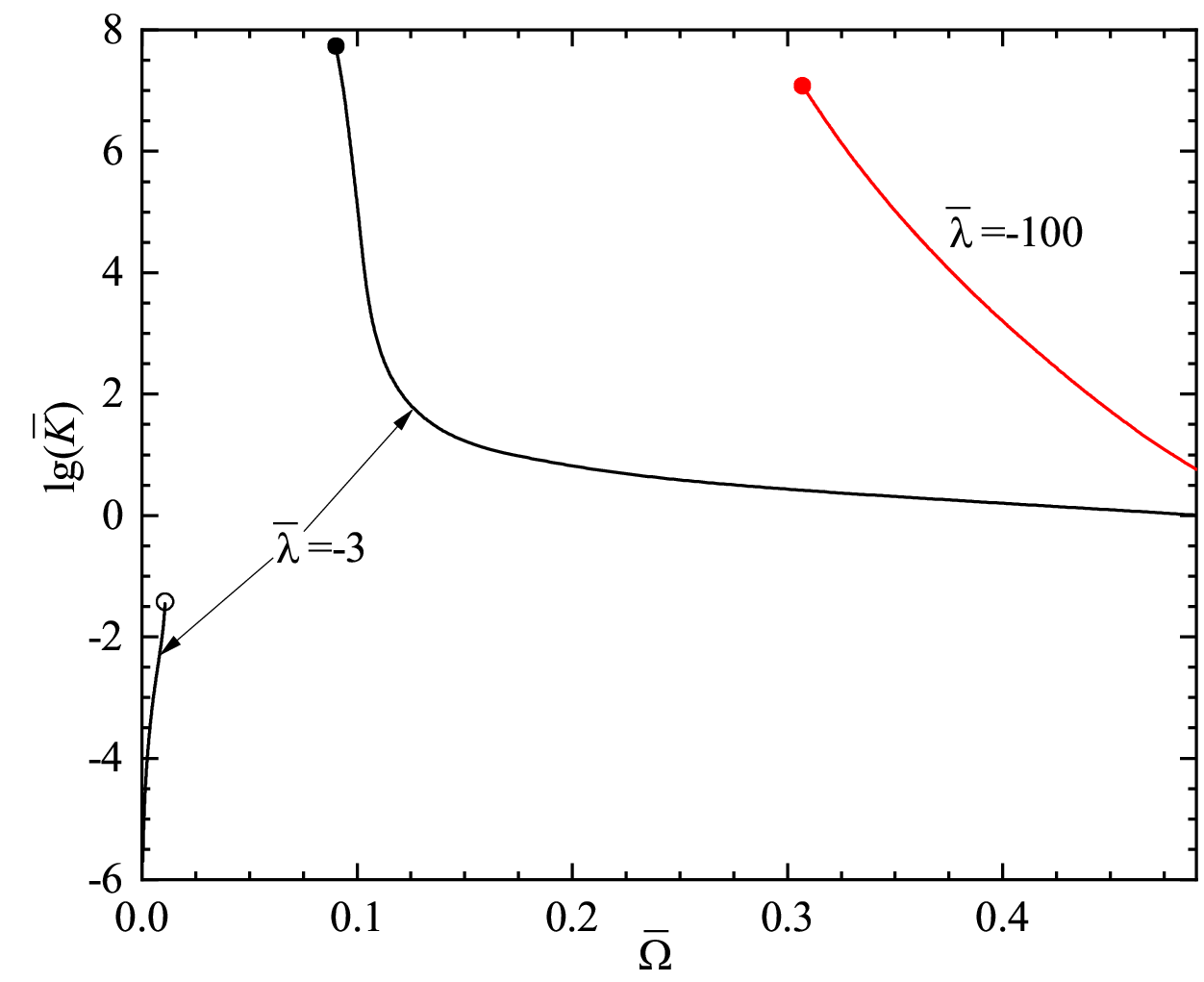}
      \end{center}
    \vspace{-.5cm}
    \caption{The minimum values of the metric function $g_{tt}\equiv e^A$ and the corresponding values of the Kretschmann scalar $\bar{K}$ in the vicinity of
     $\bar \Omega\to \bar \Omega_{1}$ (shown by solid circles) and  $\bar \Omega\to \bar \Omega_{0}$ (shown by open circles) 
    for the systems with $\bar\lambda < \bar\lambda_{\text{tv}}$. For the case of $\bar{\lambda}=-100$, the magnitude of $\bar \Omega_{0}$ is extremely small;
    this does not permit us to show the corresponding curves in the plots.
        }
    \label{fig_Kret_A}
\end{figure}

\begin{figure}[t]
    \begin{center}
        \includegraphics[width=.49\linewidth]{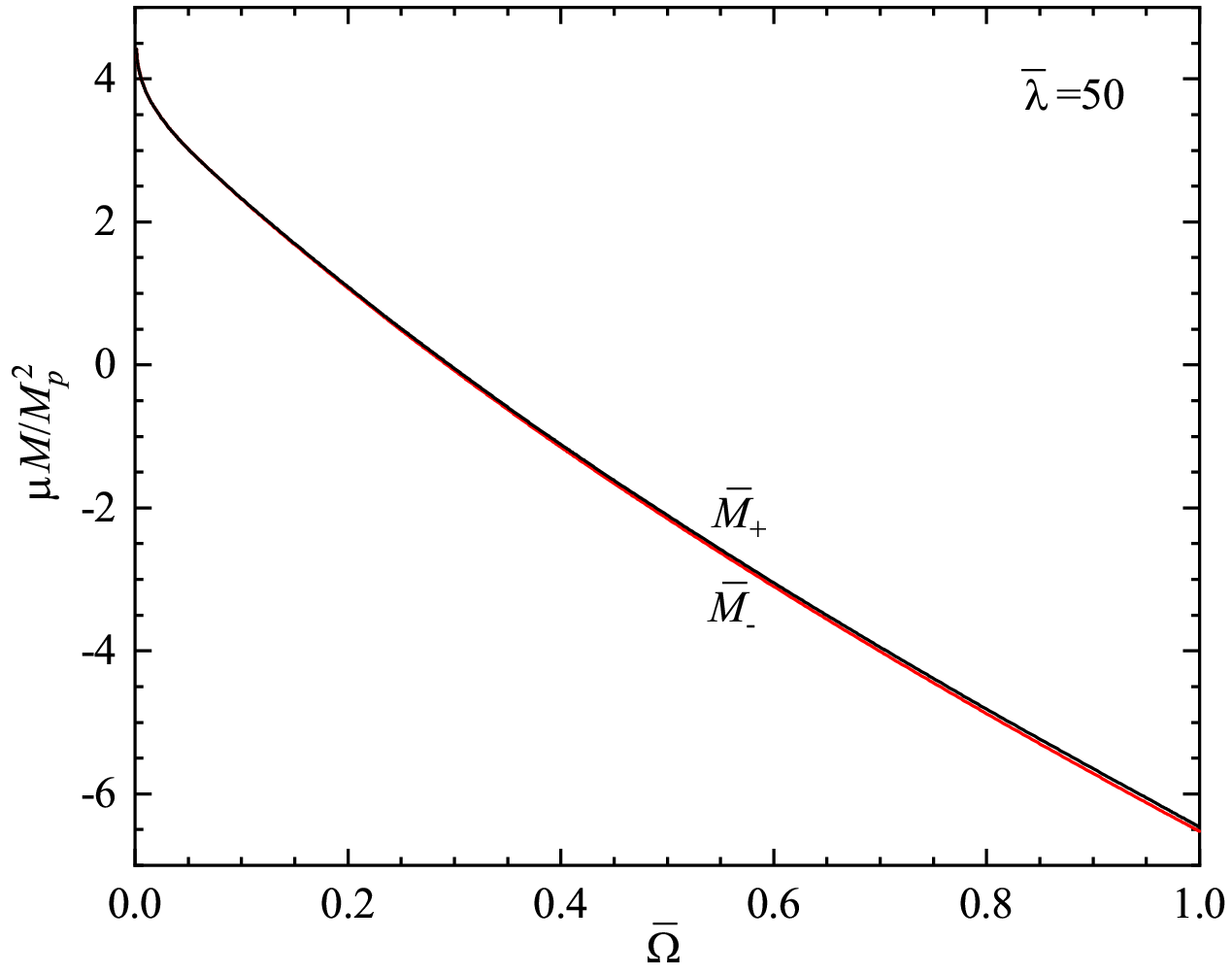}
        \includegraphics[width=.49\linewidth]{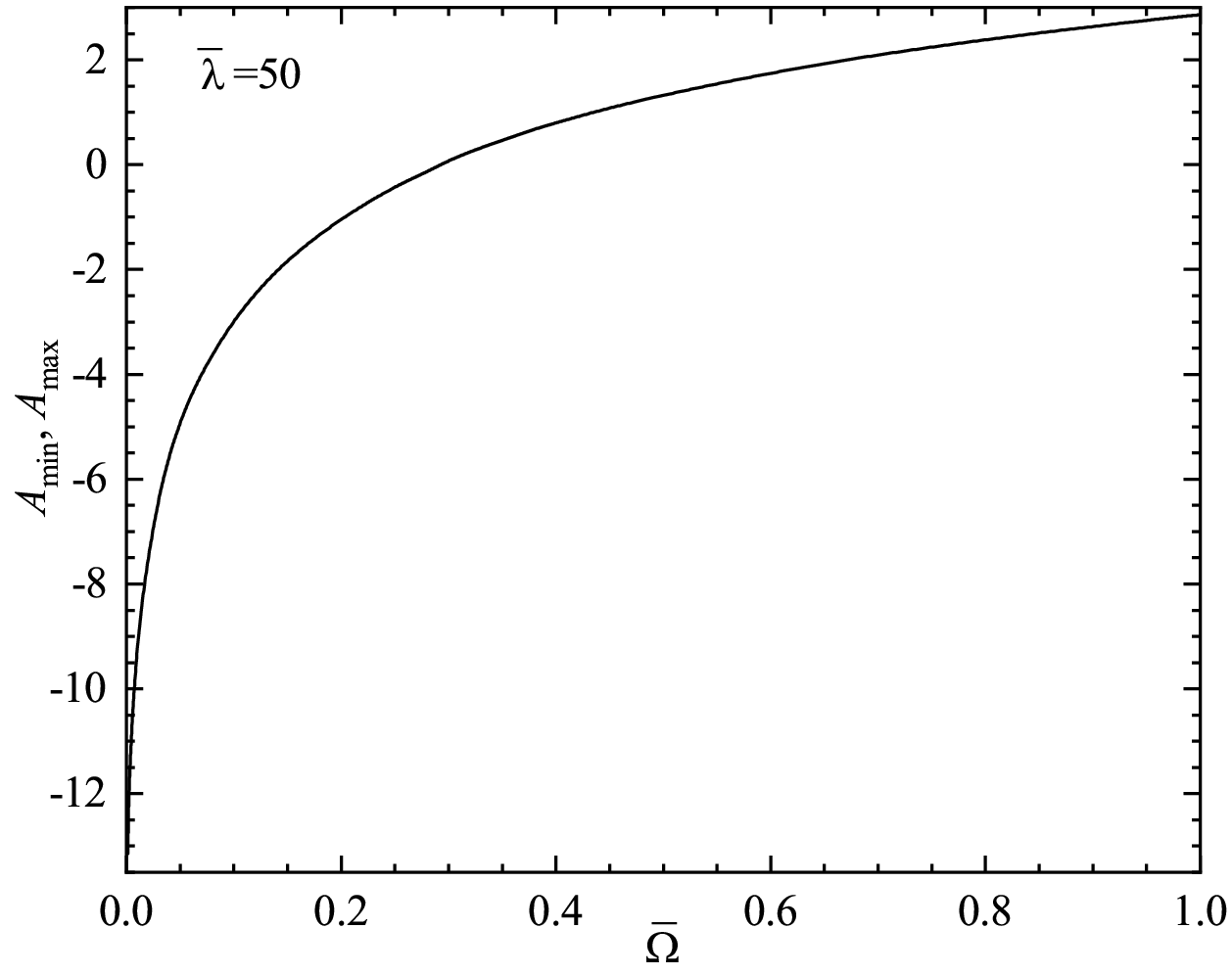}
      \end{center}
    \vspace{-.5cm}
    \caption{The case of large positive value of the nonlinearity parameter $\bar \lambda=50$.  The dimensionless Dirac-star-plus-wormhole total mass $\bar M_\pm$ (left panel) 
    and the extremal values of the metric function $A$ (right panel) as functions of the parameter $\bar\Omega$.   
    }
    \label{fig_M_A_Lambda_50}
\end{figure}

\begin{figure}[t]
    \begin{center}
        \includegraphics[width=.49\linewidth]{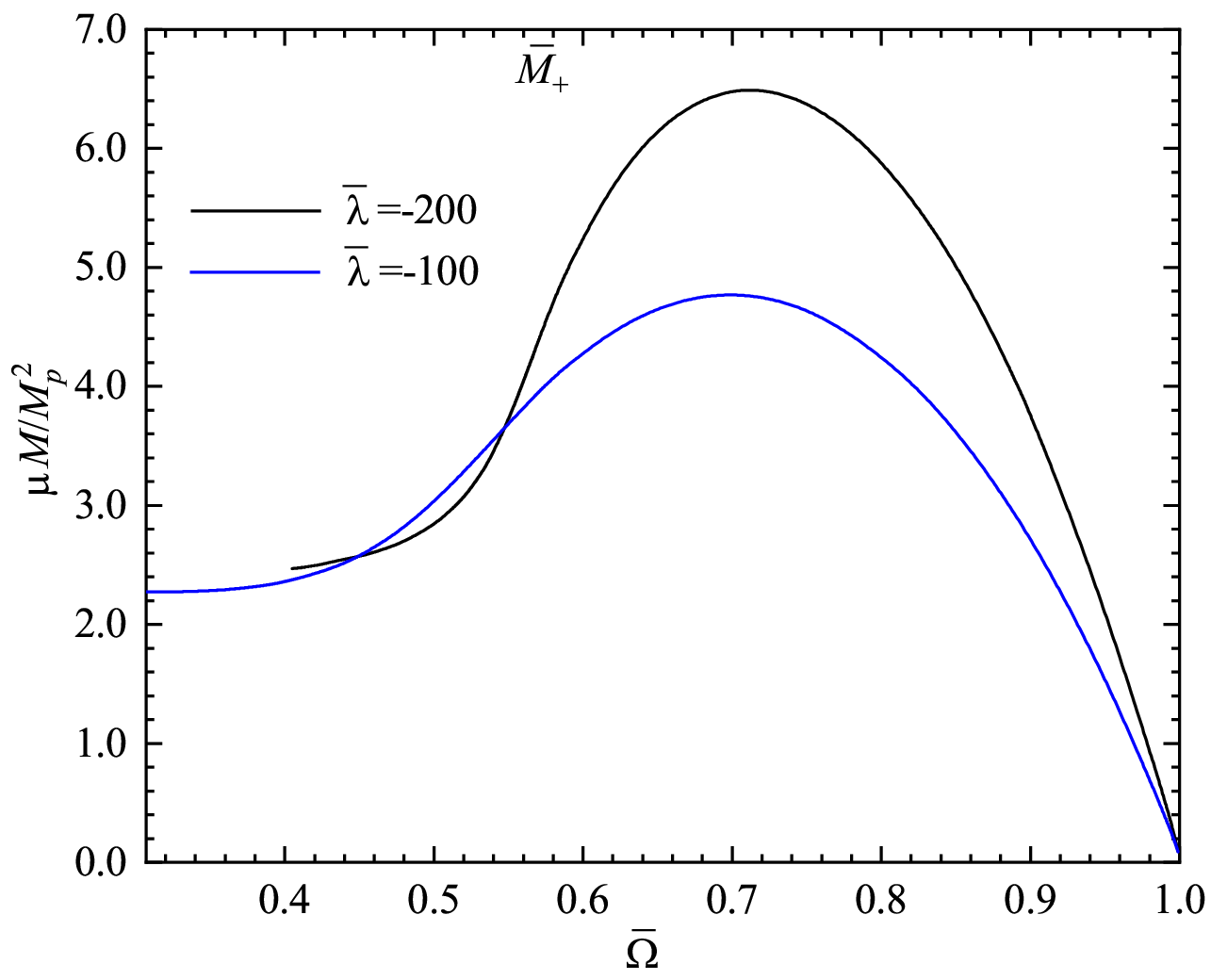}
        \includegraphics[width=.49\linewidth]{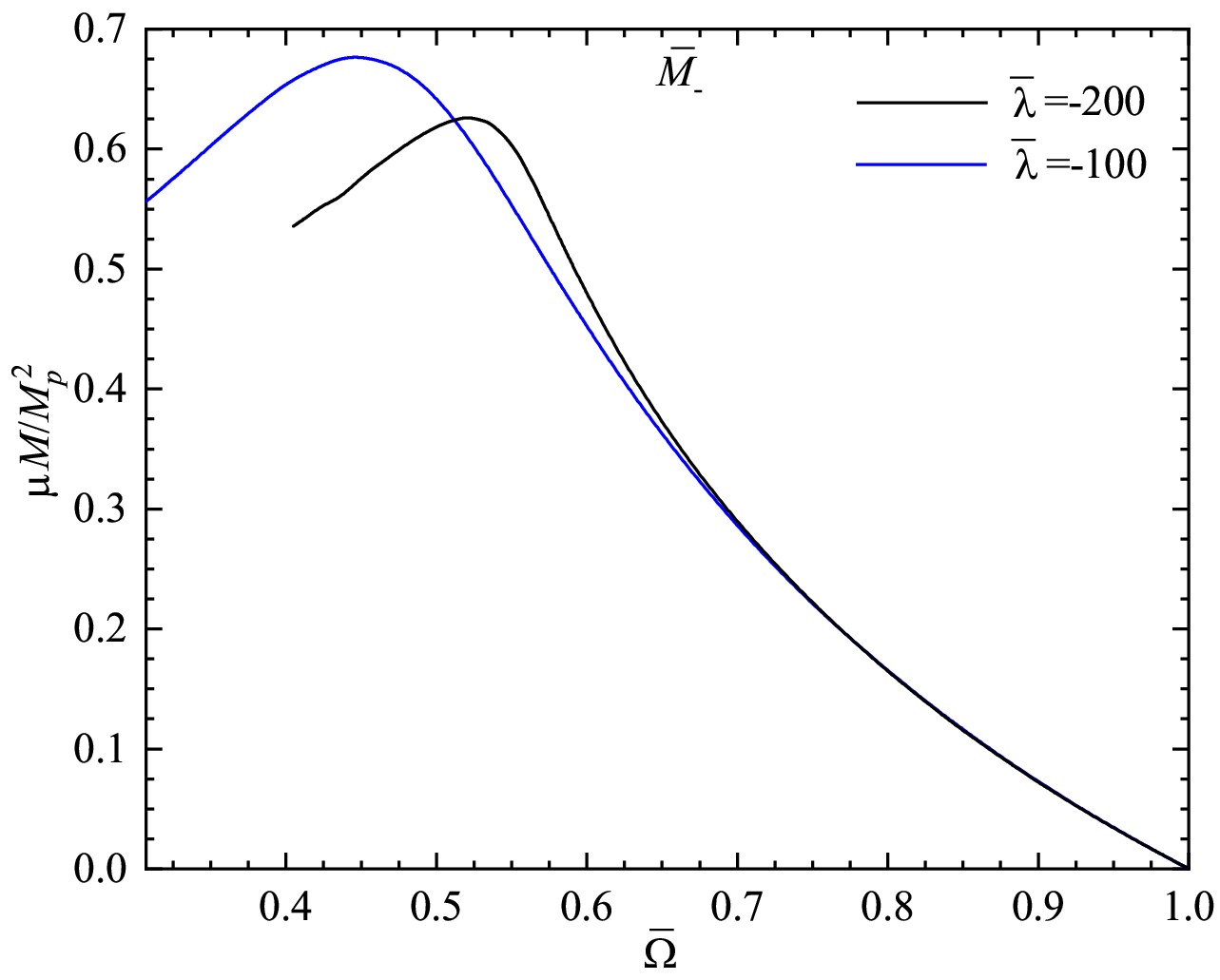}
        \includegraphics[width=.49\linewidth]{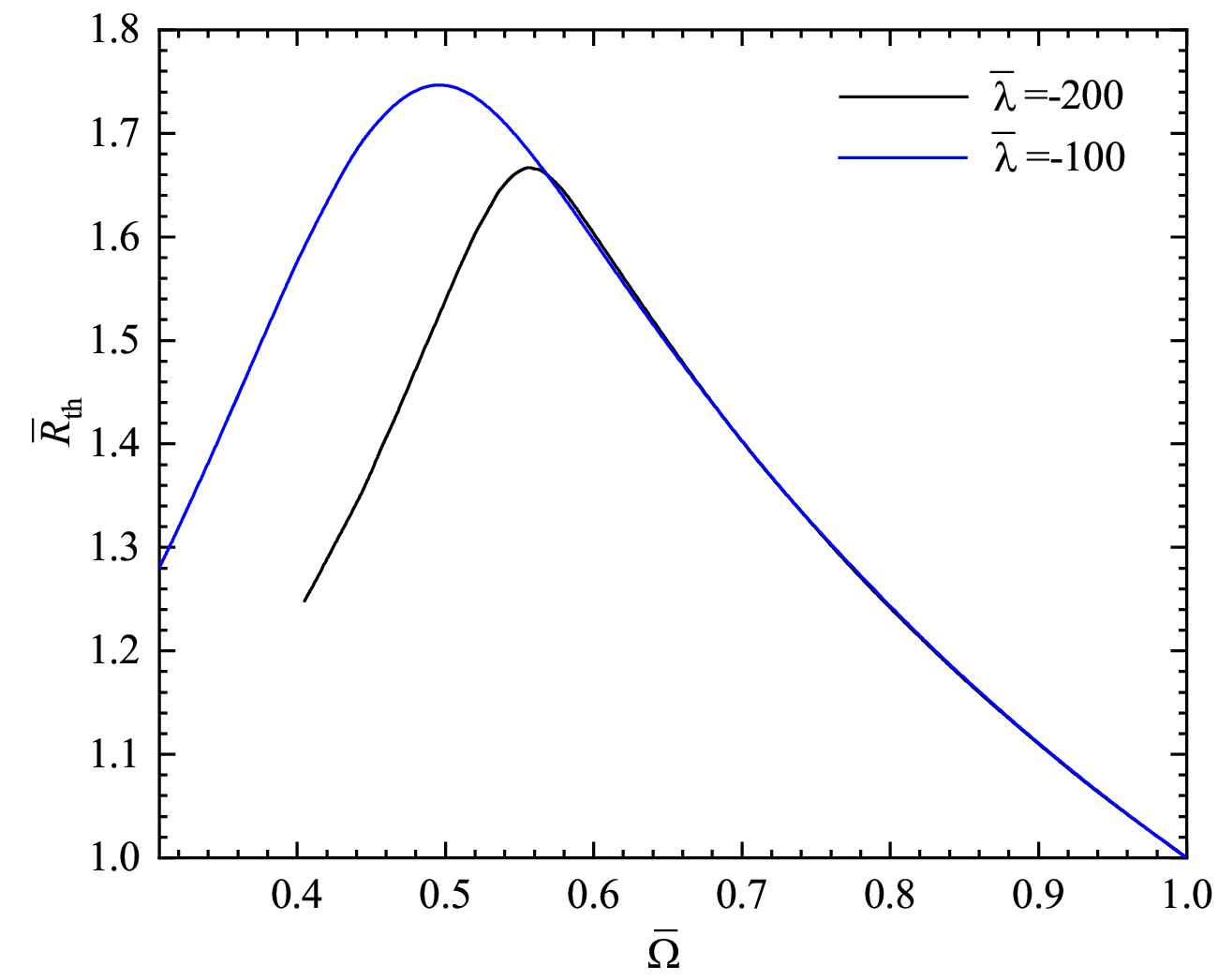}
        \includegraphics[width=.49\linewidth]{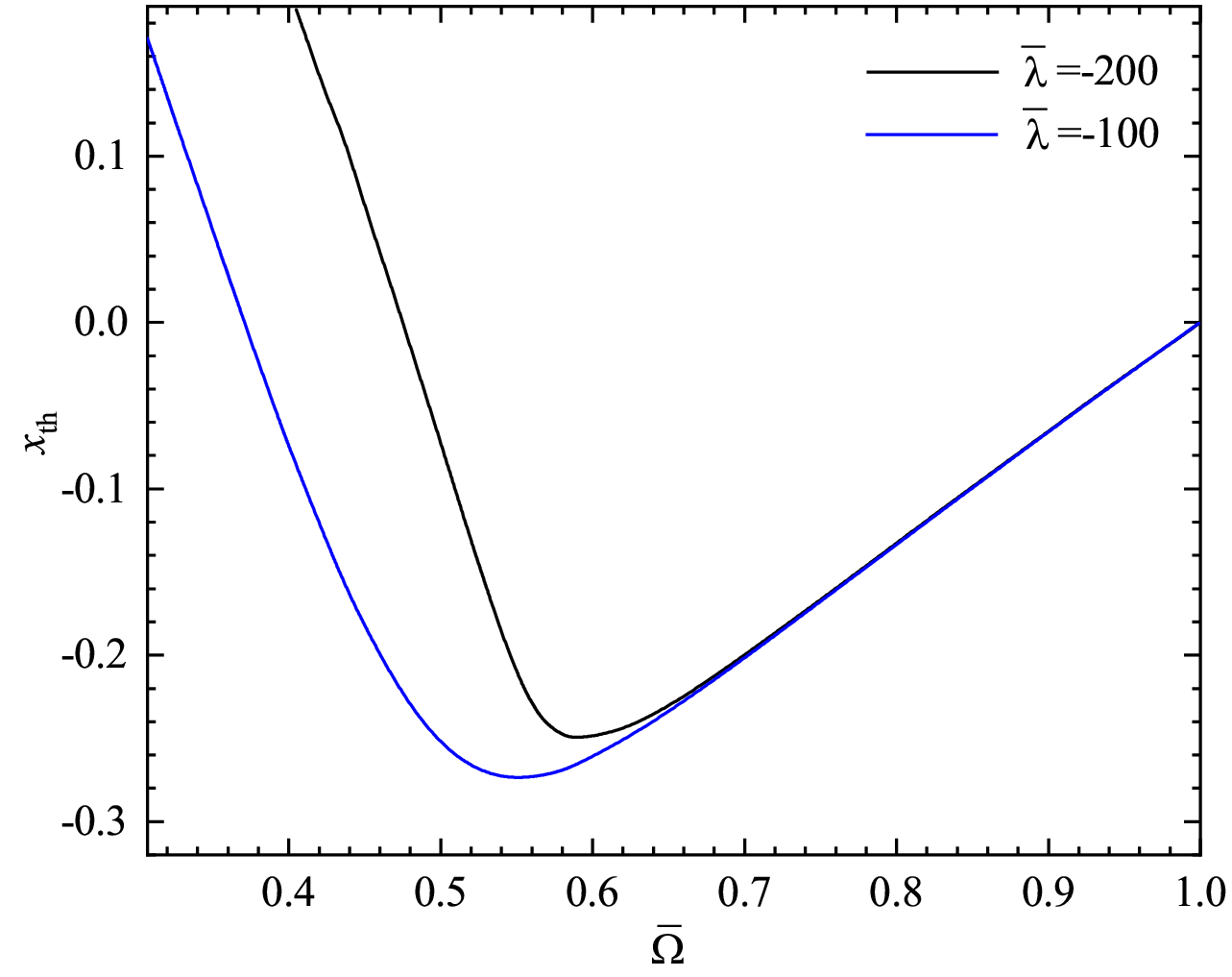}
         \end{center}
    \vspace{-.5cm}
    \caption{Top panels: the dimensionless Dirac-star-plus-wormhole total mass $\bar M_\pm$ as a function of
the parameter $\bar\Omega$ for large negative~$\bar{\lambda}$. 
Bottom panels: the dimensionless circumferential radius of the throat $\bar R_{\text{th}}$ (left panel) and the location of the throat $x_{\text{th}}$ on the $x$-axis (right panel)
as  functions of the parameter $\bar\Omega$.
}
    \label{fig_Mass_Lambda_100}
\end{figure}

\begin{figure}[h!]
    \begin{center}
        \includegraphics[width=.48\linewidth]{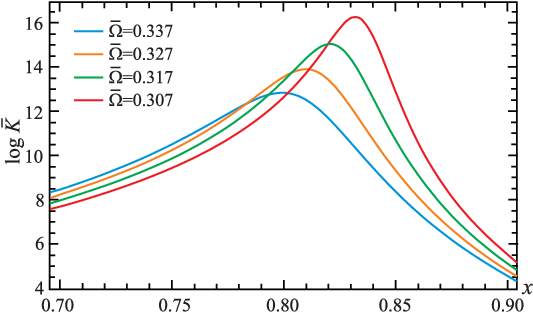}
        \includegraphics[width=.49\linewidth]{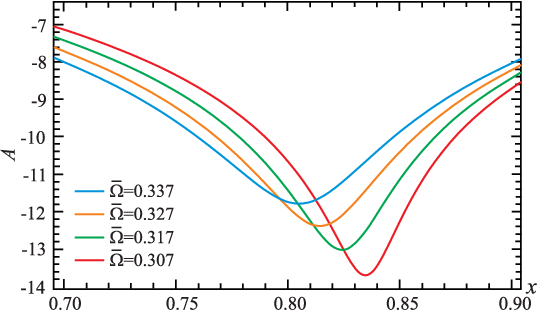}
        \end{center}
    \caption{Spatial distributions of the dimensionless Kretschmann scalar $\bar K\equiv K/\mu^4$ and the metric function $A$
   as functions of $\bar\Omega$ in the vicinity of $\bar\Omega_1$ (for $\bar \lambda=-100$).
      }
    \label{fig_Kret_A_sing}
\end{figure}

The results of numerical calculations are illustrated in Fig.~\ref{fig_plots_sols} where typical spatial distributions of the field functions for the systems with three different $\bar\lambda$ are shown.
As seen from the graphs, depending on the value of $\bar\lambda$, there can exist three types of asymmetric systems: 
(i)~the configurations possessing only one throat with the circumferential radius $\bar R_{\text{th}}$ located,  because of the asymmetry of the system, either to the left or to the right of the center
 $x=0$, depending on the value of $\bar{\Omega}$ (the case of $\bar{\lambda}=-50$); 
(ii)~the configurations only with  two throats separated by an equator with the circumferential radius $\bar R_{\text{eq}}$ (the case of $\bar{\lambda}=50$); and (iii)~the systems possessing both one and two 
throats, depending on the value of $\bar{\Omega}$  (the case of $\bar{\lambda}=-10$).

Furthermore, as seen from Fig.~\ref{fig_plots_sols}, depending on the value of the nonlinearity parameter $\bar{\lambda}$, the solutions are asymmetric to a greater or lesser extent.
Namely, for $\bar{\lambda}=50$, the solutions possess relatively small asymmetry for all values of the frequency  $\bar{\Omega}$, 
and the asymmetry becomes smaller as $\bar{\Omega}$ increases. For smaller values of $\bar{\lambda}$, the asymmetry is much stronger, and it disappears only when $\bar{\Omega}\to 1$ 
(that is, when one deals with the limiting Ellis wormhole with no spinor fields).

Fig.~\ref{fig_Mass_Omega_Lambda}  shows the dependencies of the Dirac-star-plus-wormhole total mass 
$\bar M\equiv \mu M/M_p^2$ [calculated using Eqs.~\eqref{expres_mass} or \eqref{m_current_dmls}] on $\bar \Omega$  for different
values of the nonlinearity parameter $\bar\lambda$. As in the case of a linear spinor field with $\bar\lambda =0$ considered above,
the results shown in this figure are obtained by varying the frequency $\bar \Omega$  within the limits where one can get numerical solutions. 
Namely, the calculations indicate that for $x_0=1$ there is a threshold value $\bar\lambda_{\text{tv}} =-1 $:
 for the values $\bar\lambda \geq \bar\lambda_{\text{tv}}$, it is possible to obtain solutions in all the allowed range of frequencies~$\bar \Omega$, 
 including the systems with $\bar \Omega\to 0$ and
 $\bar \Omega\to 1$ (see the left panels in Fig.~\ref{fig_Mass_Omega_Lambda}). 
 However, for $\bar\lambda < \bar\lambda_{\text{tv}}$, the behavior of the mass curves changes drastically, as illustrated in the right panels of Fig.~\ref{fig_Mass_Omega_Lambda}. 
Here, as well as in the case of a linear spinor field, there are two limiting values $\bar{\Omega}_0$ and $\bar{\Omega}_1$ which can be reached in numerical calculations
(marked by the circles in the graphs).
As $\bar{\lambda}$ increases, the difference $\left(\bar{\Omega}_{1}-\bar{\Omega}_{0}\right)$ becomes smaller and smaller, and eventually goes to zero when
$\bar{\lambda}\to -1$ (the corresponding limiting curves with $\bar{\lambda}= -1$ are shown in the right panels of 
Fig.~\ref{fig_Mass_Omega_Lambda}).

The behavior of the mass curves described above is caused by the behavior of the solutions, which is determined by the value of the nonlinearity parameter $\bar\lambda$.
Namely, for $\bar\lambda \geq \bar\lambda_{\text{tv}}$, the behavior of the metric functions $g_{tt}\equiv e^A$ and $g_{rr}\equiv B e^{-A}$ at small $\bar \Omega$ is illustrated
in Fig.~\ref{fig_gtt_grr}, using the case with $\bar\lambda=-1$ as an example. It is seen from this figure that when $\bar \Omega \to 0$ the function $g_{tt}\to 0$, while $g_{rr}$ diverges. 
In this case in all of space the Kretschmann scalar $\bar{K}$ always remains  finite, and when $\bar{\Omega}\to 0$ it tends to 0. In turn,
as $\bar{\Omega}\to 0$, the radius of the equator of such double-throat configuration increases rapidly, and one might expect that  
$\bar R_{\text{eq}}$ will eventually diverge. This will correspond to the case with a possible horizon of infinite area located at the equator
 (cold black hole).

In turn, the results of calculations for the configurations with $\bar\lambda < \bar\lambda_{\text{tv}}$ are illustrated in Fig.~\ref{fig_Kret_A},
where the dependencies of the minimum value of the metric function $g_{tt}\equiv e^A$ and the corresponding values of
the Kretschmann scalar on $\bar{\Omega}$ in the vicinity of $\bar{\Omega}_0$ and $\bar{\Omega}_1$ are shown.
It is seen from this figure that at three critical points there is the following behavior of the solutions: (i)~As $\bar{\Omega}\to 0$,
both the Kretschmann scalar and the metric function $g_{tt}$ tend to 0. At the same time, the calculations indicate that both the metric function $g_{rr}$ 
and the radius of the equator $\bar R_{\text{eq}}$ diverge rapidly in this limit. This corresponds to the fact that when $\bar{\Omega}\to 0$ 
there is a possible horizon of infinite area (cold black hole).
(ii)~When $\bar{\Omega}\to \bar{\Omega}_0$, the Kretschmann scalar is finite,  $g_{tt}\to 0$, and $g_{rr}$ is also finite. That is, a possible horizon is also present, and,
by analogy with black holes, such configurations can be referred to as black wormholes~\cite{Balakin:2007xq}.
Note that a similar situation also takes place in the case of a linear spinor field considered in Sec.~\ref{linear_field}.
(iii)~When $\bar{\Omega}\to \bar{\Omega}_1$, the Kretschmann scalar diverges,  $g_{tt}\to 0$, and $g_{rr}$ remains finite.
This corresponds to the fact that a singularity is developed in such systems, and, depending on the value of $\bar{\lambda}$, 
it is located either at the equator (for the system with $\bar{\lambda}=-3$ possessing a double throat in the vicinity of $\bar{\Omega}_1$) or at the throat 
(for the systems with $\bar{\lambda}=-10, -50$  possessing a single throat in the vicinity of $\bar{\Omega}_1$). 

Let us now consider the behavior of mass curves at large (modulus) values of $|\bar\lambda|$. The left panel of Fig.~\ref{fig_M_A_Lambda_50} 
demonstrates the dependence of the mass on $\bar \Omega$ for $\bar\lambda=50$. We see here fundamentally new features in the behavior of the curves: 
(i)~for some $\bar \Omega$, there appear negative values of the total mass~$\bar{M}$; (ii)~the masses $\bar{M}_+$ and $\bar{M}_-$ are practically equal in magnitude;
this corresponds to configurations that are practically symmetric with respect to the center (cf. Fig.~\ref{fig_plots_sols}); and
(iii)~in contrast to the configurations with smaller values of $\bar{\lambda}$ shown in Fig.~\ref{fig_Mass_Omega_Lambda},
in the limit $\bar{\Omega}\to 1$, the limiting Ellis wormholes are absent.
The corresponding solutions for such systems also have characteristic features determining such a behavior of the mass curves. The right panel of Fig.~\ref{fig_M_A_Lambda_50} 
shows the dependence of the extremal value (maximum for positive  $A$ or minimum for negative $A$) of the metric function $A$ on $\bar{\Omega}$. 
It is seen from this figure that for $\bar{\Omega}\approx 0.29$ the metric function $A$ changes the sign, and this is the reason why the total mass of the system under consideration changes the sign 
(see also Fig.~\ref{fig_plots_sols} where the spatial distribution for the metric function $g_{tt}\equiv e^A$ is shown for the case of $\bar\lambda=50$).
In turn, when $\bar \Omega \to 0$, the behavior of the solutions corresponds to the behavior of the systems with $\bar\lambda \geq \bar\lambda_{\text{tv}}$ described above.

On the other hand, as seen from Fig.~\ref{fig_Mass_Lambda_100}, for large negative $\bar\lambda$, the behavior of the mass curves becomes somewhat similar 
to that of the curves typical of gravitating systems with
trivial topology (for instance, for boson and Dirac stars, see, e.g., Refs.~\cite{Schunck:2003kk,Liebling:2012fv,Herdeiro:2017fhv,Dzhunushaliev:2018jhj}).
Here there exists a pronounced maximum of the mass, whose value either increases as  $\bar\lambda$ decreases (see the curves for $\bar{M}_+$) or, conversely, decreases
(see the curves for $\bar{M}_-$). The possibility of increasing the maximum value of the mass as $\bar\lambda$ decreases enables one to consider an astrophysically interesting situation 
where, in the limit of very large $|\bar\lambda|$, one can obtain configurations with masses and sizes typical of neutron stars (see Sec.~\ref{limit_conf}).
Furthermore, the calculations indicate that for large negative $\bar\lambda$ the systems under consideration always have only one throat,
whose circumferential radius and location on the $x$-axis as functions of the frequency  $\bar{\Omega}$ are illustrated in the bottom panels of Fig.~\ref{fig_Mass_Lambda_100}.
It is seen from these plots, in particular, that in the limit $\bar{\Omega}\to 1$ the systems tend to the limiting  Ellis
wormholes, whose throat is located at $x_{\text{th}}=0$ (symmetric wormholes), and the throat radius $\bar{R}_{\text{th}}=x_0$.

Since below we focus on astrophysically interesting systems with large negative  $\bar{\lambda}$, 
in which, as pointed out above, there develops a singularity when~$\bar \Omega\to \bar \Omega_1$, 
let us consider in more detail the evolution of
the  Kretschmann scalar and the metric function $A$ in this limit. As an example, consider the system with $\bar{\lambda}=-100$, 
for which the mass curves are given in Fig.~\ref{fig_Mass_Lambda_100}.
The rightmost point of these curves (where $\bar\Omega \to 1$) corresponds to the limiting Ellis wormhole with zero mass.
On the other hand, as $\bar\Omega$ decreases, there is the rapid increase of the Kretschmann scalar 
(see the left panel of Fig.~\ref{fig_Kret_A_sing} where the spatial distributions of $K$ in the vicinity of $\bar\Omega_1$ are shown), 
while the metric function $g_{tt}\equiv e^A$ goes to zero when $\bar \Omega\to \bar \Omega_1$ (see the right panel of Fig.~\ref{fig_Kret_A_sing}).
As pointed out above (cf. Fig.~\ref{fig_Kret_A} and the corresponding text), this implies the appearance of a spacetime singularity in the system. 
Unfortunately, the emerging singularity prevents numerical calculations in this limit.
This situation resembles the case of a boson star harbouring a wormhole considered by us earlier~\cite{Dzhunushaliev:2014bya}.

\subsection{The limiting configurations for $|\bar \lambda| \gg 1$}
\label{limit_conf}

\begin{figure}[t]
    \begin{center}
        \includegraphics[width=.49\linewidth]{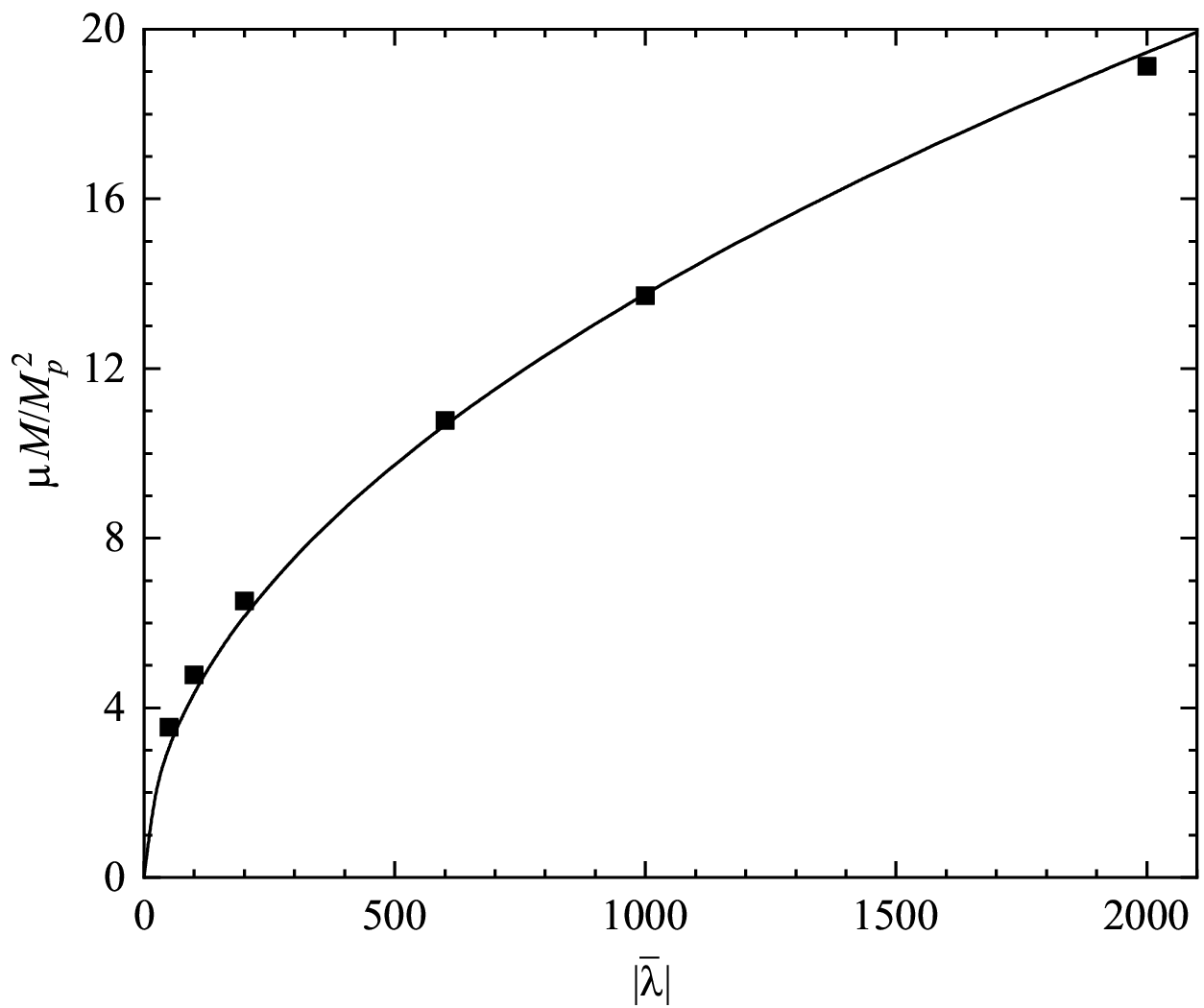}
        \includegraphics[width=.49\linewidth]{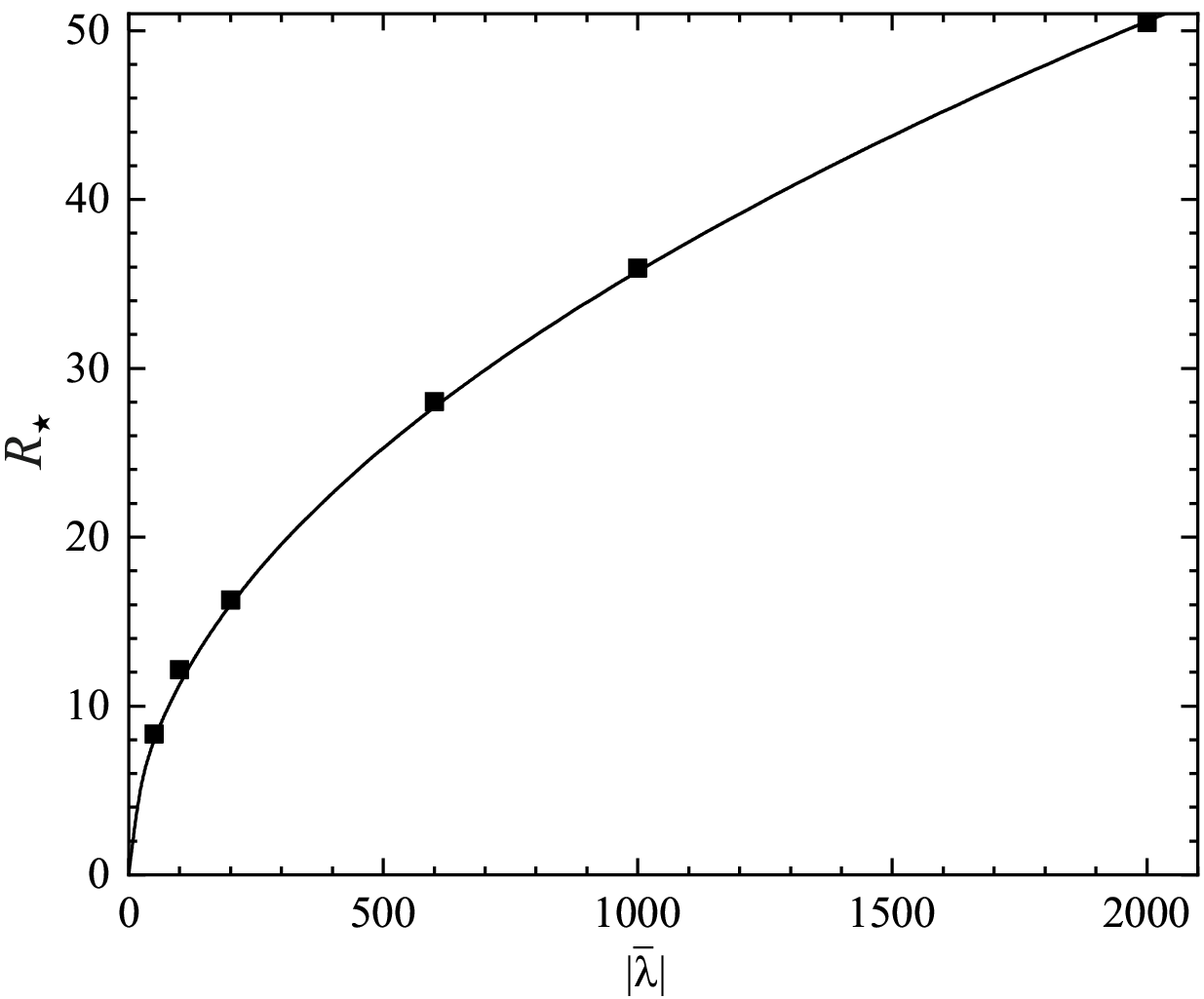}
         \end{center}
    \vspace{-.5cm}
    \caption{Maximum Dirac-star-plus-wormhole masses (left panel) and corresponding effective radii (right panel)
    as  functions of~$|\bar \lambda|$. The solid curves correspond to the asymptotic relations \eqref{M_max_approx} and \eqref{R_max_approx}, respectively.
        }
    \label{fig_mass_lambda}
\end{figure}

As shown in Ref.~\cite{Herdeiro:2017fhv},  the maximum mass of Dirac stars possessing trivial topology and supported by a linear spinor field
 is $M^{\text{max}}\approx 0.709 M_p^2/\mu$ (cf. Fig.~\ref{fig_Mass_Omega_Lambda_0}, the curve ``Dirac star'').
Then, for instance, if the mass of a spinor field is chosen to be $\mu\sim 1~\text{GeV}$, it gives the total mass $M\sim 10^{14}~\text{g}$, that is, such stars have small masses and radii of the order of
$10~\text{fm}$.
(By analogy to mini-boson stars \cite{Lee:1988av}, one can refer to such objects as mini-Dirac stars.) 
In turn, the inclusion of a nonlinear spinor field may result in a considerable increase in masses and sizes of configurations 
with trivial topology~\cite{Dzhunushaliev:2018jhj,Dzhunushaliev:2019kiy,Dzhunushaliev:2019uft}.
As one can see from the results obtained above, a similar situation occurs for the systems with nontrivial topology considered:
the use of negative values
of the parameter $\bar \lambda$ leads to increasing the maximum mass. In this connection, from the point of view of possible astrophysical applications,
it is interesting to employ negative values of $\bar \lambda$ to obtain larger masses. Numerical calculations indicate that as $|\bar \lambda|$ increases,
there is a considerable growth in maximum masses of the systems under consideration
(see the top left panel of Fig.~\ref{fig_Mass_Lambda_100} for $\bar{M}_+$). For clarity,
in Fig.~\ref{fig_mass_lambda}, we have extended the range of $\bar \lambda$ to larger (modulus) values and
plotted the dependence of the maximum mass
 $M^{\text{max}}$ on $|\bar \lambda|$. In this figure, the solid line corresponds to the interpolation formula
 \begin{equation}
\label{M_max_approx}
	M^{\text{max}}\approx 0.435 \sqrt{|\bar \lambda|}M_p^2/\mu,
\end{equation}
which holds asymptotically for $|\bar \lambda|\gg 1$.

For our purpose, it is useful to recast the above formula in a form which is more appropriate from the astrophysical point of view. 
To do this, we reparameterize the families of gravitational equilibria by a single dimensionless quantity
$\bar \lambda =\lambda M_p^2 c/4\pi \hbar^3$, as was done in Ref.~\cite{Dzhunushaliev:2018jhj} (here we temporarily restore $c$ and $\hbar$). 
Consistent with the dimensions of
 $[\lambda]=\text{erg cm}^3$, one can assume that its characteristic value is
  $\lambda \sim  \tilde \lambda \,\mu c^2 \lambda_c^3$, where $\lambda_c$ is the Compton wavelength and
  the dimensionless quantity $\tilde \lambda \sim 1$.
  Then the dependence of the maximum mass on
  $|\bar \lambda|$  in the limit  $|\bar \lambda|\gg 1$  given by Eq.~\eqref{M_max_approx} can be represented as
\begin{equation}
\label{M_max_approx_Sun}
	M^{\text{max}}\approx   0.2\, \sqrt{|\tilde \lambda|}M_{\odot}\left(\frac{\text{GeV}}{\mu}\right)^2.
\end{equation}
For the typical mass of a fermion $\mu\sim 1~\text{GeV}$, the above mass is comparable to the Chandrasekhar mass.
In this respect the dependence of the maximum mass of the Dirac-star-plus-wormhole configurations on the coupling constant~$\lambda$
is similar to that of boson stars~\cite{Colpi:1986ye, Mielke:2000mh} or Dirac stars with trivial topology~\cite{Dzhunushaliev:2018jhj,Dzhunushaliev:2019kiy,Dzhunushaliev:2019uft}.

Let us now turn to the radius of the configurations under consideration. Since such systems do not possess a sharp surface, 
their sizes are not uniquely defined. Here we adopt the following definition for the effective radius that 
is rather insensitive to the various definitions employed~\cite{Kleihaus:2011sx,Dzhunushaliev:2014bya}
 \begin{equation}
\label{R_max_def}
R_{\star}=\frac{\int_{x_{\text{max}}}^{\infty}\sqrt{-g} j^t R(r) dr}{\int_{x_{\text{max}}}^{\infty}\sqrt{-g} j^t dr} ,
\end{equation}
where $R$ is defined by Eq.~\eqref{circ_radius} and the expression for the timelike component of the four-current
$j^t$ is given below Eq.~\eqref{BE_gen}. In turn, the definition of  the lower limit in the integrals can be found above Eq.~\eqref{part_num}.
Using the expression~\eqref{R_max_def}, in the right panel of Fig.~\ref{fig_mass_lambda},
we have plotted the dependence of the effective radius of configurations with maximum masses (given in the left panel of Fig.~\ref{fig_mass_lambda}) on $|\bar \lambda|$. 
In this figure, the solid line corresponds to the interpolation formula
 \begin{equation}
\label{R_max_approx}
	R_{\star}^{\text{max}}\approx 1.13 \sqrt{|\bar \lambda|}/\mu\approx 0.77 \sqrt{|\tilde \lambda|}\left(\frac{\text{GeV}}{\mu}\right)^2\text{km},
\end{equation}
which holds asymptotically for $|\bar \lambda|\gg 1$.
For the typical mass of a fermion $\mu\sim 1~\text{GeV}$, the above expression gives radii of the order of
kilometers. In combination with the masses of the order of the Chandrasekhar mass [see Eq.~\eqref{M_max_approx_Sun}], this
corresponds to characteristics typical of neutron stars.

\section{Stability}
\label{stability}

Let us now briefly address the question of stability of the systems under consideration. Our previous investigations of mixed systems consisting of a wormhole 
threaded either by ordinary neutron matter~\cite{Dzhunushaliev:2013lna,Dzhunushaliev:2014mza} or by usual (non-ghost) scalar field~\cite{Dzhunushaliev:2014bya}
showed that such systems are unstable with respect to radial linear perturbations. 
As in the case of static Ellis wormholes supported by a massless ghost scalar field~\cite{Gonzalez:2008wd,Gonzalez:2008xk,Blazquez-Salcedo:2018ipc}
and of wormholes consisting of scalar fields with a potential~\cite{Dzhunushaliev:2013lna,Dzhunushaliev:2017syc},
this radial instability is caused by the presence of a ghost scalar field in the static mixed configurations. 
By continuity, it is therefore expected that the mixed configurations studied in the present paper, in which the neutron matter or the usual scalar field 
are replaced by a spinor field, will also be dynamically unstable. 
However, the inclusion of an asymmetric spinor field tremendously complicates a linear stability analysis.

Nevertheless, even without performing an ambitious stability analysis, it is clear that, for stable configurations, the binding energy (BE)
must necessarily be positive (energy stability). In the case of the mixed Dirac-star-plus-wormhole
systems under consideration, this simple stability criterion is also applicable (see the corresponding discussion of this question in Ref.~\cite{Dzhunushaliev:2012ke} where a mixed 
star-plus-wormhole system supported by neutron matter and a massless ghost scalar field has been considered). 
Consistent with this, let us  calculate
 the BE,
which is defined as the difference between the energy of $N_f$ free particles, ${\cal E}_f=N_f \mu$, and the total energy of the system, ${\cal E}_t=M$,
that is, 
 \begin{equation}
\label{BE_gen}
\text{BE}={\cal E}_f-{\cal E}_t .
\end{equation}
Here the total particle number
 $N_f$ is equal to the Noether charge $Q$ of the system, which is defined via the timelike component of the four-current $j^\alpha=\bar \psi \gamma^\alpha \psi$
as
$
Q=\int\sqrt{-g} j^t d^3 x ,
$
where in our case $\sqrt{-g}j^t = B^{3/2}e^{-3 A/2}\left(r^2+r_0^2\right) \sin{\theta} \left(\psi^\dag \psi\right)$.
For symmetric systems, the lower limit in the above integral is the point $r=0$ where a throat or an equator are located.

However, for the asymmetric systems under consideration, the situation is more complicated, 
since in such systems there are either one throat or two throats and equator located in general asymmetrically with respect to the center.
In this case the lower limit in the above integral can be determined as a point $r=r_{\text{max}}$ where the maximum of the charge density of the spinor field 
 $j^t=\bar{\psi}\gamma^0\psi=e^{-A/2}\left(\psi^\dag \psi\right)$ is located.
In terms of the dimensionless variables \eqref{dmls_var}, we then have for the Noether charge:
\begin{equation}
\label{part_num}
N_{f\pm}=Q_\pm=\pm \left(\frac{M_p}{\mu}\right)^2\int_{x_{\text{max}}}^{\pm\infty} B^{3/2}e^{-3 A/2}\left(x^2+x_0^2\right)\left(\bar u^2+\bar v^2\right) dx.
\end{equation}

Alternatively, one can determine the particle number for asymmetric systems using the trick suggested in Ref.~\cite{Hoffmann:2018oml}. 
To do this, one introduces a fictitious electromagnetic field, 
such that the particle number can be read off asymptotically as the electric charge associated with such field. In doing so, it is assumed that such electromagnetic field has no backreaction
on the spinor and gravitational fields but is sourced by the spinor field. The Lagrangian of the  electromagnetic field can be chosen in the form
$$
L_{\text{em}}=-\frac{1}{4}F_{\mu\nu}F^{\mu\nu} ,
$$
where $F_{\mu\nu}=\partial_\mu A_\nu-\partial_\nu A_\mu$ is the electromagnetic field tensor. Then the corresponding Maxwell equations are
 $$
\frac{1}{\sqrt{-g}}\frac{\partial}{\partial x^\nu}\left(\sqrt{-g}F^{\mu\nu}\right)=-j^\mu\equiv-\bar{\psi}\gamma^\mu\psi .
$$
 As {\it Ansatz} for the four-potential $A_\mu$, one takes here $A_\mu=\left\{a(r),0,0,0\right\}$; this corresponds to the presence of the radial electric field.
 Substitution of this {\it Ansatz} in the Maxwell equations yields the corresponding second-order ordinary differential equation for the electric potential $a$, 
 which is solved as a two-point boundary value problem with the boundary conditions in the form $a\left(r\to\pm\infty\right)=0$.
In turn, the asymptotic ($r\to \pm \infty$) behavior of the potential  $a$ has the form
$$
 a\approx \pm \frac{1}{4\pi}\frac{Q_\pm}{r} \quad \Rightarrow \quad \bar{a}\approx \pm \frac{\bar{Q}_\pm}{x} .
$$
Here the second expression is written in terms of the dimensionless variables~\eqref{dmls_var}, and we also introduced new dimensionless quantities 
 $\bar{a}=4\pi G \mu \,a$ and $\bar{Q}=\left(\mu/M_p\right)^2 Q$. The charge $Q$ appearing here is the Noether charge corresponding to the particle number $N_f$ of the spinor field.
 Then the corresponding BE can be calculated using formula~\eqref{BE_gen}.
 Note that in numerical calculations it is more suitable to compute the charge using the derivative of~$a$:   $Q_\pm\to \mp \left(M_p/\mu\right)^2 x^2 \bar{a}^\prime $ as $x\to \pm \infty$.
It can be directly verified that this charge $Q$ coincides with that calculated using formula~\eqref{part_num}. 

A necessary condition for the energy stability of the configurations under consideration is the positivity of the BE. In this connection, in
the mass curves given in Fig.~\ref{fig_Mass_Omega_Lambda_0}, we have plotted the parts of the curves where the BE is negative by dashed lines.
In turn, as seen from Fig.~\ref{fig_Mass_Lambda_100}, the systems with $|\bar{\lambda}|\gg 1$, which are interesting from the astrophysical point of view, 
are energetically stable.

\section{Conclusion}
\label{concl}

In the present paper, we have continued our previous investigations of the mixed systems consisting of a wormhole threaded by various types of ordinary (non-ghost) matter:
the neutron-star-plus-wormhole configurations~\cite{Dzhunushaliev:2011xx,Dzhunushaliev:2012ke,Dzhunushaliev:2013lna,Dzhunushaliev:2014mza,Aringazin:2014rva,Dzhunushaliev:2015sla,Dzhunushaliev:2016ylj,Dzhunushaliev:2022elv} 
and wormholes threaded by bosonic matter~\cite{Dzhunushaliev:2014bya}.
In order to have a nontrivial wormhole-type topology in the system,
we have used here a massless ghost scalar field. In such a wormhole, we have added 
two spinor fields having opposite spins;  this enabled us
to get a diagonal energy-momentum tensor suitable for describing 
 spherically symmetric systems. For such a mixed system, we have constructed families of regular asymptotically
flat solutions for explicitly time-dependent spinor fields, oscillating with a frequency~$\Omega$. 
The resulting mixed configurations are asymmetric with respect to the center, in contrast to all mixed systems considered by us earlier. 
Because of the asymmetry,  masses and sizes of the configurations observed at the two asymptotic ends of the wormhole (that is, when $r\to \pm \infty$) 
may differ considerably.
It is shown that, depending on the value of the nonlinearity parameter~$\bar\lambda$, 
these solutions describe configurations possessing both positive and negative ADM masses. It is also demonstrated that energetically stable configurations
with positive masses may have interesting astrophysical applications when in the limit of large values
 $|\bar{\lambda}|\gg 1$ such systems possess masses and sizes comparable to those of ordinary neutron stars.
 This enables one to use such solutions for a description of compact gravitating objects~-- Dirac stars with nontrivial topology.

Notice the most interesting features of the systems under consideration: 
\begin{itemize}
\item[(i)] In the case of a linear spinor field ($\bar\lambda=0$), a key factor determining the properties of the configurations 
is the value of the throat parameter $x_0$. For small magnitudes of this parameter, the range of allowed values of the frequency $\bar{\Omega}$,
for which there exist regular solutions with finite energy, is divided into two subranges 
$0<\bar{\Omega}<\bar{\Omega}_0$ and $\bar{\Omega}_1<\bar{\Omega}<1$ with $\bar{\Omega}_0<\bar{\Omega}_1$, 
where the magnitude of the limiting values $\bar{\Omega}_0$ and $\bar{\Omega}_1$ is determined by the value of $x_0$; 
the systems with $\bar{\Omega}\to \bar{\Omega}_1$ are singular (the Kretschmann scalar diverges), while the configurations with $\bar{\Omega}\to \bar{\Omega}_0$
have a possible regular horizon (black wormholes). 
For larger values of $x_0$ regular solutions do exist in all the range $0<\bar{\Omega}<1$.
Whatever the value $x_0$, as $\bar{\Omega}\to 0$, the systems have two throats, and the radius of their equator increases rapidly
(the appearance of a possible horizon with infinite area~-- a cold black hole).
\item[(ii)] In the case of a nonlinear spinor field, depending on the value of the nonlinearity parameter $\bar\lambda$, 
there can exist three types of asymmetric systems: 
(a)~the configurations possessing only one throat with the circumferential radius $\bar R_{\text{th}}$ located,  
because of the asymmetry of the system, either to the left or to the right of the center $x=0$, depending on the values of $\bar{\Omega}$; 
(b)~the configurations only with  two throats separated by an equator; 
(c)~the systems possessing both one and two throats, depending on the value of~$\bar{\Omega}$.
\item[(iii)] There is some threshold value $\bar\lambda_{\text{tv}}$ separating two families of solutions: (a)~for $\bar\lambda\geq\bar\lambda_{\text{tv}}$,
the solutions are regular (the Kretschmann scalar is finite in all of space) in all the range of frequencies  $0<\bar{\Omega}<1$,  but when $\bar{\Omega}\to 0$ the metric function $g_{tt}\to 0$,
while $g_{rr}$ and $\bar R_{\text{eq}}$ diverge (cold black hole); (b)~for $\bar\lambda < \bar\lambda_{\text{tv}}$, as in the case of a linear spinor field, there are two limiting values
 $\bar{\Omega}_0$ and $\bar{\Omega}_1$, and when $\bar{\Omega}\to \bar{\Omega}_0$ the Kretschmann scalar is finite (a possible regular horizon~-- configurations of the type ``black wormholes''), while when
$\bar{\Omega}\to \bar{\Omega}_1$ it diverges rapidly (the appearance of a spacetime singularity). In turn, as  $\bar{\Omega}\to 0$, the systems of the type ``cold black holes'' do arise. 
\item[(iv)] For large positive $\bar{\lambda}$, the configurations may possess a negative total mass in some range of frequencies $\bar{\Omega}$.
\item[(v)] For large negative  $\bar{\lambda}$, the total mass and the radius of the systems under consideration increase proportionally to $\sqrt{|\bar{\lambda}|}$,
see Eqs.~\eqref{M_max_approx} and~\eqref{R_max_approx}, respectively.  In the limiting case $|\bar{\lambda}|\gg 1$, one can get objects with the mass of the order of $M_{\odot}$ and
with the size of the order of kilometers. 
\end{itemize}

A point that still remains unclear is the presence of a gap in the frequency range between $\bar{\Omega}_0$ and  $\bar{\Omega}_1$ where we could not find solutions. 
One can assume that this is related to the appearance of
singularity when $\bar{\Omega}\to \bar{\Omega}_1$ from the right, while when  $\bar{\Omega}\to \bar{\Omega}_0$ from the left, the Kretschmann scalar remains finite
and $g_{tt}\to 0$. This enables us to assume that in the latter case there appears a possible event horizon, but, using the static coordinates in the form~\eqref{metric}, it is impossible to cross the horizon when
 $\bar{\Omega}_0<\bar{\Omega}<\bar{\Omega}_1$. Perhaps this situation is similar to that of occurring for a Reissner-Nordstr\"{o}m black hole:
 for some values of the mass $m$ and of the charge $q$, there is a naked singularity, while for another values there is one or two event horizons.
 In our case, it is reasonable to suggest that the presence of a possible event horizon and of a singularity depends on specific values of the parameters $\bar{\Omega}$ and $x_0$.
 Unfortunately, a direct verification of the existence of the event horizon in numerical calculations faces obvious technical problems.

\section*{Acknowledgements}

We gratefully acknowledge support provided by the program
No.~BR24992891 (Integrated research in nuclear, radiation physics and engineering, high energy physics and cosmology for the development of competitive technologies)
of the Committee of Science of the Ministry of Science and Higher Education of the Republic of Kazakhstan.

\end{document}